\documentclass[a4paper,11pt]{article}
\pdfoutput=1 

\usepackage{jcappub} 

\usepackage{rotating}
\usepackage{tablefootnote}

\def \hubbleUnit {km s$^{-1}$Mpc$^{-1}$}
\def \HFilterRed {85.0^{+6.0}_{-6.3}}
\def \HFilterBlue {88.5^{+7.5}_{-7.4}}
\def \HFilterBoth {86.6^{+6.8}_{-6.9}}

\def \HBothFlatSource {84.3^{+6.7}_{-7.0}}
\def \HBothAniFlatBeta {75.7^{+8.3}_{-7.8}}
\def \HBothAniFlatSource {74.5^{+8.0}_{-7.8}}

\def \alphaRed {5.82 \cdot 10^{-4}}
\def \alphaBlue {8.20 \cdot 10^{-4}}
\def \alphaBoth {6.22 \cdot 10^{-4}}
\def \alphaBothAni {6.30 \cdot 10^{-4}}

\def \interceptRed {-3.18 \cdot 10^{-3}}
\def \interceptBlue {-4.70 \cdot 10^{-3}}
\def \interceptBoth {-3.43 \cdot 10^{-3}}
\def \interceptBothAni {-3.64 \cdot 10^{-3}}

\def \sigmaRed {9.78 \cdot 10^{-5}}
\def \sigmaBlue {1.32 \cdot 10^{-4}}
\def \sigmaBoth {1.04 \cdot 10^{-4}}
\def \sigmaBothAni {1.41 \cdot 10^{-4}}

\def \skewnessRed {-0.307}
\def \skewnessBlue {-0.333}
\def \skewnessBoth {-0.307}
\def \skewnessBothAni {-0.089}

\def \timeDelayError {1.6\%}
\def \losError {4.7\%}
\def \lensError {2.8\%}
\def \kinematicsError {7.5\% }
\def \totalErrorGaussian {9.4\% }
\def \totalErrorFull {7.9\%}

\def \HubbleSuyua {78.7^{+4.3}_{-4.5}}
\def \HubbleSuyub {80.0^{+4.5}_{-4.7}}

\title{The mass-sheet degeneracy and time-delay cosmography: 
Analysis of the strong lens RXJ1131-1231}

\author[a]{Simon Birrer}
\emailAdd{simon.birrer@phys.ethz.ch}
\author[a]{Adam Amara}
\emailAdd{adam.amara@phys.ethz.ch}
\author[a]{Alexandre Refregier}
\emailAdd{alexandre.refregier@phys.ethz.ch}

\affiliation[a]{Institute for Astronomy, Department of Physics, ETH Zurich\\ Wolfgang-Pauli-Strasse 27, 8093, Zurich, Switzerland}

\abstract{
We present extended modelling of the strong lens system RXJ1131-1231 with archival data in two HST bands in combination with existing line-of-sight contribution and velocity dispersion estimates. Our focus is on source size and its influence on time-delay cosmography. We therefore examine the impact of mass-sheet degeneracy and especially the degeneracy pointed out by Schneider \& Sluse (2013) \cite{Schneider:2013p8677} using the source reconstruction scale. We also extend on previous work by further exploring the effects of priors on the kinematics of the lens and the external convergence in the environment of the lensing system. Our results coming from RXJ1131-1231 are given in a simple analytic form so that they can be easily combined with constraints coming from other cosmological probes. We find that the choice of priors on lens model parameters and source size are subdominant for the statistical errors for $H_0$ measurements of this systems. The choice of prior for the source is sub-dominant at present (2\% uncertainty on $H_0$) but may be relevant for future studies. More importantly, we find that the priors on the kinematic anisotropy of the lens galaxy have a significant impact on our cosmological inference. When incorporating all the above modeling uncertainties, we find  $H_0 = \HFilterBoth$ \hubbleUnit, when using kinematic priors similar to other studies. When we use a different kinematic prior motivated by Barnab\`e et al. (2012) \cite{Barnabe:2012p11353} but covering the same anisotropic range, we find $H_0 = \HBothAniFlatSource$ \hubbleUnit. This means that the choice of kinematic modeling and priors have a significant impact on cosmographic inferences. The way forward is either to get better velocity dispersion measures which would down weight the impact of the priors or to construct physically motivated priors for the velocity dispersion model.
}

\arxivnumber{1511.03662}

\begin{document}
\maketitle
\flushbottom

\section{Introduction}
\noindent Strong lensing systems and the time delays between different images of the same background source can provide information about angular diameter distance relations (see \cite{Refsdal:1964p10230} and review of \cite{Blandford:1992p10378} for the early work). Cosmographic analyses rely on measurements of time delay \citep[see e.g.,][and the COSMOGRAIL collaboration]{Courbin:2005p10859, Eigenbrod:2005p10477, Courbin:2011p10868, Tewes:2012p10123, RathnaKumar:2013p12138, Tewes:2013p5278}\footnote{www.cosmograil.org} and estimates of the line-of-sight structure and lensing potential. This cosmography technique has been applied to determine the Hubble parameter $H_0$ using different strong lens systems \citep[see e.g.][]{Schneider:1995p10412, Kochanek:1996p10235, Schechter:1997p10303,Koopmans:2003p10513, Wucknitz:2004p10515, York:2005p10546, Jakobsson:2005p10577, Vuissoz:2007p10579, Paraficz:2009p10585, Fadely:2010p10613, Suyu:2010p4938, Suyu:2013p4952, Suyu:2014p8316} and also by applying statistics to multiple systems \citep[see e.g.][]{Saha:2006p9215, Oguri:2007p10451, Coles:2008p10703}. In the past, some of the measurements have produced a wide range of results for $H_0$ \citep[e.g. see section 8.2 of][]{Suyu:2010p4938}. One concern has been to evaluate the impact of potential systematic errors. In particular, the mass-sheet degeneracy (MSD) \cite{Falco:1985p8873} and related degeneracies that cause biases due to model assumptions \citep[e.g.][]{Schneider:1995p10412, Saha:2000p9143, Wucknitz:2002p10026, Liesenborgs:2012p12099, Schneider:2014p12100} need special consideration. For instance, this has been illustrated by \cite{Schneider:2013p8677} where they show that assuming a power-law lens model can cause significant biasing of results.

In this paper, we introduce a new treatment of the MSD and source reconstruction for cosmographic analyses. This approach integrates information coming from imaging, velocity dispersion, external convergence and time delay measurements. For the choice of data and the parameterization of the lens we follow the work of \cite{Suyu:2013p4952}, and we infer the values of the parameters using our recent framework presented in \cite{Birrer:2015p11550}. In our framework we reconstruct the source using shapelet basis sets. This allows us to explicitly set an overall scale for the reconstruction. We will show that this enables us to better disentangle the effects coming from source structure and MSD. This then makes it simpler to robustly combine the information coming from the different data sets.

The paper is organized as follow: In Section \ref{sec:theory} we briefly review the principles of time delay cosmography. Section \ref{sec:data} presents the data used in this work. Section \ref{sec:lens_modeling} describes the details of the lens modeling, including kinematics, likelihood analysis and the source reconstruction technique of \cite{Birrer:2015p11550}. In Section \ref{sec:degeneracies} we show that the use of this reconstruction technique turns out to be well designed for mapping out the MSD. Section \ref{sec:posterior_sampling} describes the combined likelihood analysis and posterior sampling. Section \ref{sec:cosmological_inference} discuss the cosmological constraints in terms of angular diameter relations and cosmological parameters. In Section \ref{sec:comparison}, we compare our results to others. We summarize our conclusions in Section \ref{sec:summary}.

\section{Theory} \label{sec:theory}
Gravitational lensing is caused by deflection of light by matter. In this section, we review the principles of gravitational lensing and time delay cosmography and introduce our conventions.

\subsection{Lensing formalism}
The lensing potential $\psi(\vec{\theta})$ at an angular position $\vec{\theta} = (\theta_1, \theta_2)$ is given by
\begin{equation} \label{eqn:lens_pot}
  \psi(\vec{\theta}) = \frac{1}{\pi} \int d^2 \vec{\theta'} \kappa(\vec{\theta'}) \ln |\vec{\theta} - \vec{\theta'}|
\end{equation}
where $\kappa$ is the convergence and is given by
\begin{equation}
  \kappa(\vec{\theta}) = \frac{\Sigma(D_{\text{d}} \vec{\theta})}{\Sigma_{\text{crit}}}
\end{equation}
with
\begin{equation} \label{eqn:Sigma_crit}
  \Sigma_{\text{crit}} = \frac{c^2D_{\text{s}}}{4\pi G D_{\text{d}} D_{\text{ds}}}
\end{equation}
is the critical density and $\Sigma(D_{\text{d}} \vec{\theta})$ is the physical projected surface mass density. $D_{\text{d}}$, $D_{\text{s}}$ and $D_{\text{ds}}$ are the angular diameter distances from the observer to the lens, to the source and from the lens to the source \footnote{$D_{\text{ds}}$ is \textit{not} the subtraction $D_{\text{d}}-D_{\text{s}}$. In a flat universe: $D_{\text{ds}}= \frac{1}{1 + z_s} (M_d - M_s)$, where $M$ is the transverse co-moving distance.}, respectively. The deflection angle is $\vec{\alpha}(\vec{\theta}) = \vec{\nabla} \psi(\vec{\theta})$ and the lens equation, which describes the mapping from the source plane $\vec{\beta}= (\beta_1, \beta_2)$ to the image plane $\vec{\theta}$ is given by
\begin{equation}
  \vec{\beta} = \vec{\theta} - \vec{\alpha}(\vec{\theta}).
\end{equation}
The convergence $\kappa(\vec{\theta})$ can also be written as
\begin{equation}
  \kappa(\vec{\theta}) = \frac{1}{2} \nabla^2 \psi(\vec{\theta}).
\end{equation}

\subsection{Time delays}
The Fermat potential is defined as
\begin{equation}
  \phi(\vec{\theta}, \vec{\beta}) \equiv \left[ \frac{(\vec{\theta} - \vec{\beta})^2}{2} - \psi(\vec{\theta}) \right].
\end{equation}
The excess time delay of an image at $\vec{\theta}$ with corresponding source position $\vec{\beta}$ is
\begin{equation}
  t(\vec{\theta}, \vec{\beta}) = \frac{D_{\Delta t}}{c} \phi(\vec{\theta}, \vec{\beta})
\end{equation} 
where
\begin{equation} \label{eqn:D_t_definition}
  D_{\Delta t} \equiv (1 + z_{\text{d}}) \frac{D_{\text{d}}D_{\text{s}}}{D_{\text{ds}}}
\end{equation}
is referred as the time delay distance. The relative time delay difference $\Delta t_{ij}$ between two images positioned at $\vec{\theta}_i$ and $\vec{\theta}_j$, the actual observable, is then given by
\begin{equation} \label{eqn:delta_t_definition}
  \Delta t_{ij} = t_i(\vec{\theta_i}, \vec{\beta}) - t_j(\vec{\theta}_j, \vec{\beta}).
\end{equation}
Line-of-sight (LOS) structures external to the lens also affect the observed time delay distance through additional focusing or de-focusing of the light rays. We parameterize the LOS structure with a single constant mass sheet parameter $\kappa_{\text{ext}}$, the external convergence. The actual time delay distance $D_{\Delta t}$ relates to the one inferred by ignoring the external LOS structure by
\begin{equation}
  D_{\Delta t} = \frac{D_{\Delta t}^{\text{model}} }{1 - \kappa_{\text{ext}}}.
\end{equation}

\section{RXJ1131-1231 system} \label{sec:data}
\begin{figure*}
  \centering
  \includegraphics[angle=0, width=140mm]{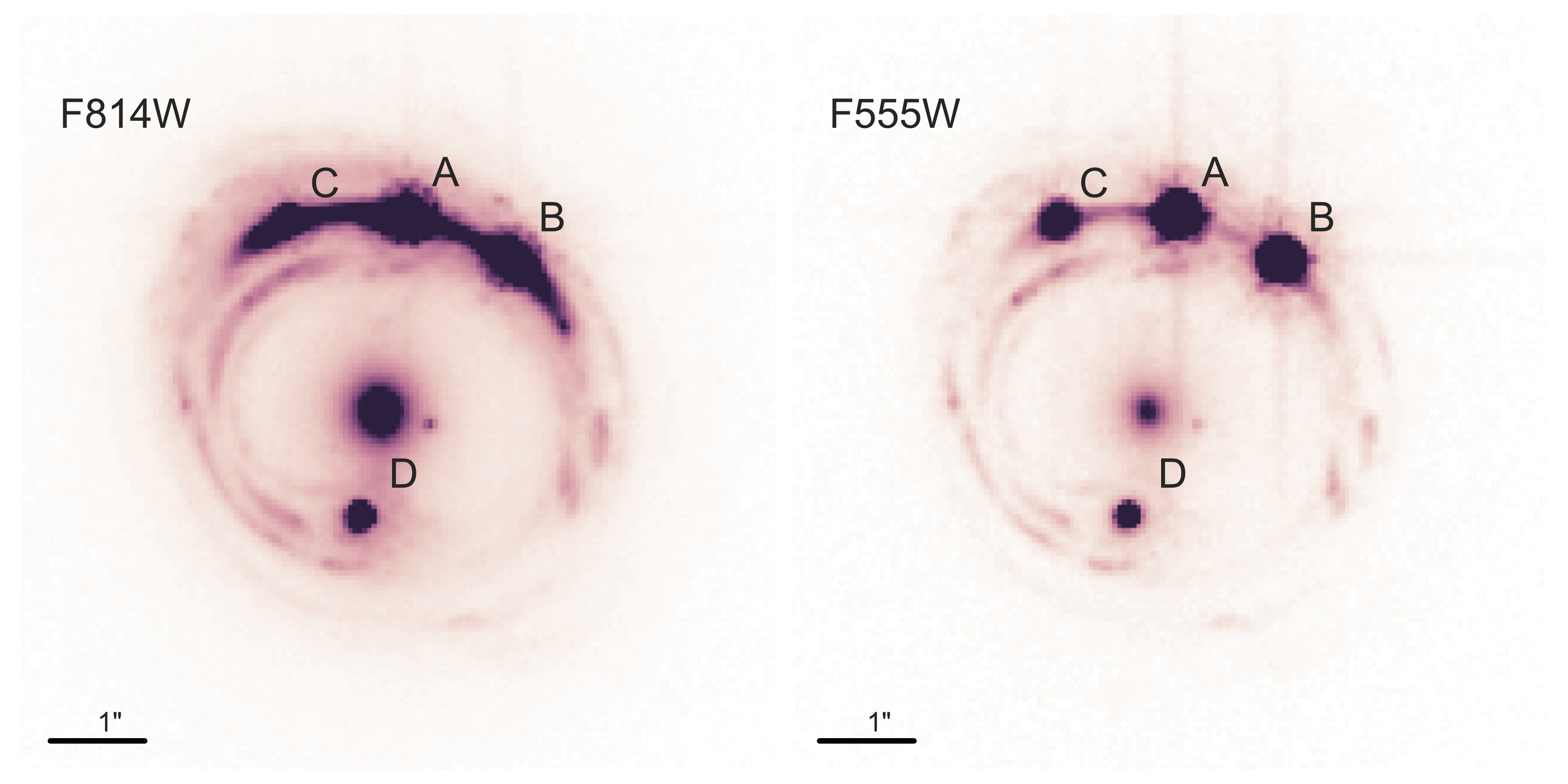}
  \caption{HST ACS WFC1 images in filters F814W (left) and F555W (right). The F814W filter has more high signal-to-noise pixels than the F555W filter. In the F555W filter, the substructure in the Einstein ring and the diffraction spikes of the quasar images are more prominent. The letters A,B,C,D indicate the quasar images for the time delay differences.}
\label{fig:original_images}
\end{figure*}

The quadrupole lens system RXJ1131-1231 (Figure \ref{fig:original_images}) was discovered by \cite{Sluse:2003p8680} and the redshift of the lens $z_l = 0.295$ and of the background quasar source $z_s = 0.658$ was determined spectroscopically by \cite{Sluse:2003p8680}. The lens was modeled extensively by \cite{Claeskens:2006p8872, Brewer:2008p8808, Suyu:2013p4952, Birrer:2015p11550, Chen:2016p11356} with single band images. We use the archival HST ACS WFC1 images in filter F814W and F555W (GO 9744; PI: Kochanek). The filter F814W was also used for lens modeling in \cite{Suyu:2013p4952}, \cite{Suyu:2014p8316} and \cite{Birrer:2015p11550}. We make use of the \texttt{MultiDrizzle} product from the HST archive. We use a 160$^2$ pixel image centered at the lens position with pixel scale 0.05". This corresponds to a FOV of 8".

For the analysis in this work, we take the time delay measurements and uncertainties from \cite{Tewes:2013p5281}, namely $\Delta t_{AB} = 0.7 \pm 1.4$ days, $\Delta t_{CB} = -0.4 \pm 2.0$ days, and $\Delta t_{DB} = 91.4 \pm 1.5$ days, where $[A,B,C,D]$ represent the quasar images in Figure \ref{fig:original_images}. This data was used in \cite{Suyu:2013p4952}, where they also measure the LOS velocity dispersion of $\sigma_v = 323 \pm 20$ km s$^{-1}$, that we use in our analysis. 

For the external convergence $\kappa_{\text{ext}}$, we take the estimate of \cite{Suyu:2013p4952} based on relative galaxy counts in the field \citep[][]{Fassnacht:2011p11092} and their modeled external shear component compared with ray tracing of the Millennium Simulation (see their Figure 6). As their probability density function for $\kappa_{\text{ext}}$ is not given in a parameterized form, we use an approximation of their PDF in the form of a skewed normal distribution with mean $\mu_{\kappa}= 0.1$, standard deviation $\sigma_{\kappa} = 0.042$ and skewness $\gamma_{\kappa} = 0.8$. This function is illustrated in Figure \ref{fig:kappa_pdf} and described in Appendix \ref{app:skew_normal_distribution}.

\begin{figure}
  \centering
  \includegraphics[angle=0, width=80mm]{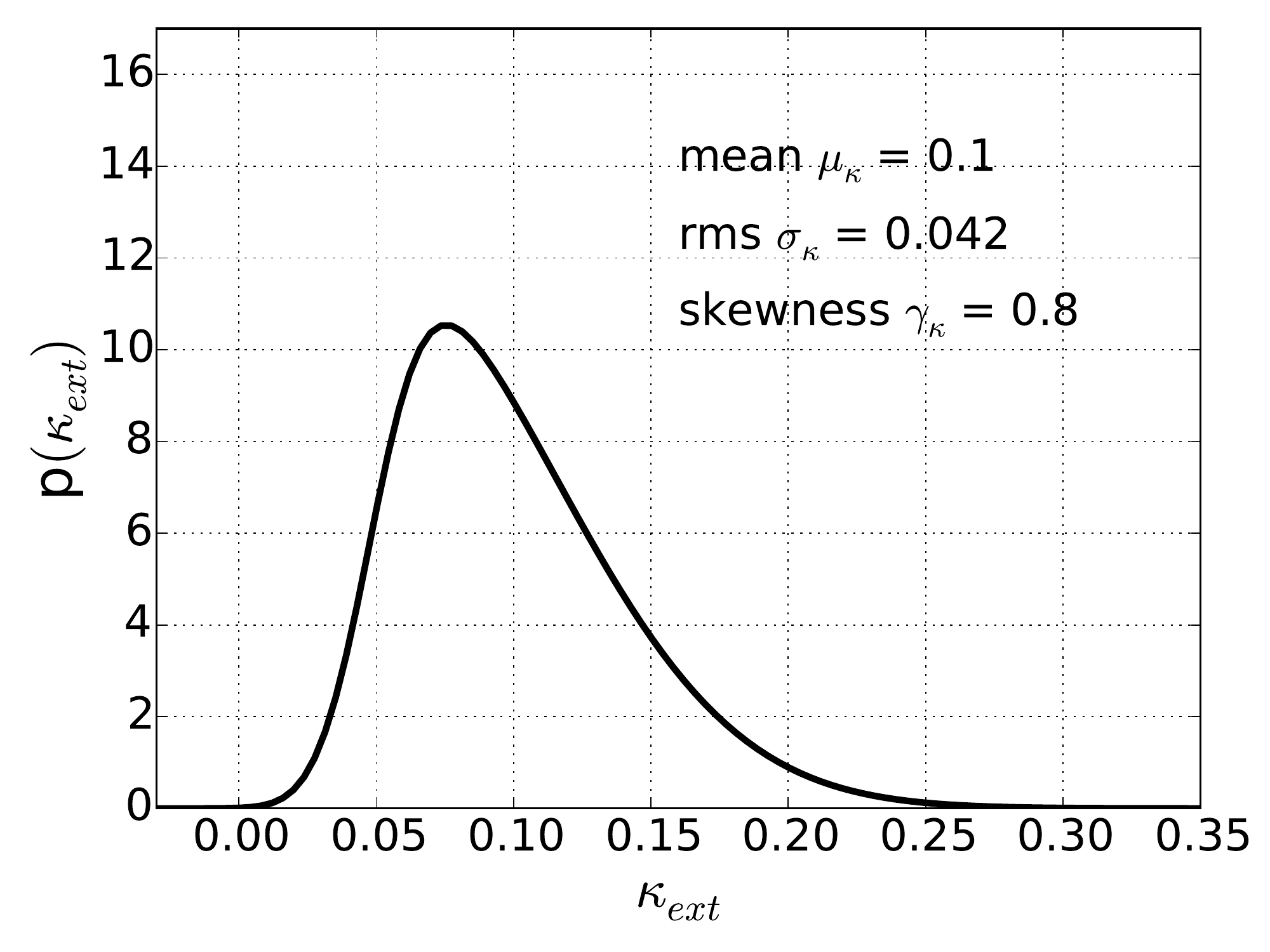}
  \caption{Probability density function of the external convergence in the form of a skewed normal distribution. The parameters chosen are designed to match well the probability density function quoted in \cite{Suyu:2013p4952} (their Figure 6).}
\label{fig:kappa_pdf}
\end{figure}

\section{Lens modeling} \label{sec:lens_modeling}
In this section, we present the parameterization of the lens model, the lens light description, the source reconstruction technique, PSF modeling, the modeling of the lens kinematics and the likelihood analysis.

\subsection{Lens model parameterization} \label{sec:lensmodel_param}
For the lens model, we use:
\begin{enumerate}
  \item An elliptical power-law mass distribution parameterized as
\begin{equation} \label{eqn:lens_model_ellipse}
  \kappa_{\text{lens}}(\theta_1, \theta_2) = \frac{3 - \gamma'}{2} \left( \frac{\theta_E}{\sqrt{q\theta_1^2 + \theta_2^2/q}} \right)^{\gamma'-1}
\end{equation}
where $\theta_E$ is the Einstein radius, $q$ is the ellipticity and $\gamma'$ is the radial power-law slope.
  \item A second spherical isothermal profile (Equation \ref{eqn:lens_model_ellipse} with fixed $\gamma'=2$ and $q=1$) centered at the position of the visible companion of the lens galaxy about 0.6 arc seconds away from the center.
  \item A constant external shear yielding a potential parameterized in polar coordinates $(\theta, \varphi)$ given by
\begin{equation}
  \psi_{\text{ext}}(\theta, \varphi) = \frac{1}{2} \gamma_{\text{ext}}\theta^2 \cos 2(\varphi-\phi_{\text{ext}})
\end{equation}
with $\gamma_{\text{ext}}$ is the shear strength and $\phi_{\text{ext}}$ is the shear angle.
\end{enumerate}

\subsection{Lens light parameterization} \label{sec:lens_light_param}
The light distribution of the lens is modeled in a parameterized form. We use the same profiles as \cite{Suyu:2013p4952}, namely two elliptical S\'ersic profiles \citep[][]{Sersic:1968p10728} with common centroid for the central elliptical galaxy and an additional spherical S\'ersic profile for the companion galaxy. The intensity profile is parameterized as
\begin{equation}
  I(\theta_1,\theta_2) = A \exp \left[ -k \left(  \left( \frac{\sqrt{\theta_1^2 + \theta_2^2/q_{\text{L}}^2}}{\theta_{\text{eff}}} \right)^{1/n_{\text{sersic}}}-1\right) \right]
\end{equation}
where $A$ is the amplitude, $k$ is a constant such that $\theta_{\text{eff}}$ is the effective half-light radius, $q_{\text{L}}$ is the axis ratio and $n_{\text{sersic}}$ is the S\'ersic index. We use the value of half-light radius $\theta_{\text{eff}}$ as the effective radius in the kinematics modeling of Section \ref{sec:stellar_kinematics}.

\subsection{Source surface brightness reconstruction} \label{sec:source_reconstruction}
We use the source reconstruction method presented in \cite{Birrer:2015p11550} based on shapelet basis functions introduced by \cite{Refregier:2003p8153}. To apply this method, three choices have to be made. (1) The shapelet center position, which we fixed to quasar source position. The determination of the quasar source position is explained in detail in Section 4.2 of \cite{Birrer:2015p11550}. (2) The width of the shapelet basis function $\beta$ (see Section \ref{sec:degeneracies} for its impact). (3) The maximal order $n_{\text{max}}$ of the shapelet polynomials. We set $n_{\text{max}}=30$ for modeling and parameter inference. With this, most of the features in the extended source can be modeled. Given these three choices, one can reconstruct the angular scales between $\beta/\sqrt{n_{\text{max}}+1}$ and $\beta\sqrt{n_{\text{max}}+1}$ around the center of the shapelet in the source plane.

\subsection{PSF modeling} \label{sec:psf_modeling}
We use four bright stars in the same ACS image to model the PSF. After normalizing for flux, we apply a sub-pixel shift to recenter the stars and then stack. When comparing the individual star images and the stack, we see significant variations that we need to consider in our analysis. To do this by measuring the scatter for each pixel and assume that the scatter in high signal-to-noise pixels is due to a model error that we quantify as a fraction of the flux. This leads to an additional error term, beyond the Poisson and background contribution, that is important close to the center of the bright point sources (see Section \ref{sec:likelihood_analysis}). For the quasar point sources, we use a cutout of the PSF of 111$^2$ pixels to cover most of the diffraction spikes. For the extended surface brightness we apply a PSF-convolution kernel of 21$^2$ pixels.

\subsection{Stellar kinematics} \label{sec:stellar_kinematics}
We follow the analysis of \cite{Suyu:2010p4938} for the modeling of the stellar velocity dispersion. The mass profile is assumed to be a spherical symmetric power-law in the form of
\begin{equation}
  \rho_{\text{local}}(r) = \rho_0 \left( \frac{r_0}{r} \right)^{\gamma'}
\end{equation}
where $\rho_0$ is the density at radius $r_0$ and $\gamma'$ is a power-law slope of the mass profile (the same $\gamma'$ as for the lens model in Equation \ref{eqn:lens_model_ellipse}). The normalization of the mass profile can be expressed in terms of the lensing quantities as
\begin{equation}
  \rho_0 r_0^{\gamma'} = (\kappa_{\text{ext}} - 1) \Sigma_{\text{crit}} \theta_{\text{E}}^{\gamma' -1} D_{\text{d}}^{\gamma' -1} \frac{\Gamma \left( \frac{\gamma'}{2}  \right)}{\pi^{1/2} \Gamma \left( \frac{\gamma' -3}{2}  \right)}.
\end{equation}
where $\kappa_{\text{ext}}$ is the external convergence, $\Sigma_{\text{crit}}$ is the critical projected density, $\theta_{\text{E}}$ is the Einstein radius, $D_{\text{d}}$ is the angular diameter distance from the observer to the lens and $\Gamma$ is the Gamma function.
The estimation of the projected velocity dispersion along the line of sight requires a description of the anisotropic velocity component split in radial and tangential component
\begin{equation}
  \beta_{\text{ani}} \equiv 1 -  \frac{\sigma_z^2}{\sigma_r^2}.
\end{equation}
Massive elliptical galaxies are assumed to have isotropic stellar motions in the center of the galaxy ($\beta_{\text{ani}}=0$) and radial motions in the outskirts ($\beta_{\text{ani}}=1$). A simplified description of the transition can be made with an anisotropy radius parameterization $r_{\text{ani}}$ defining $\beta_{\text{ani}}$ as a function of radius $r$ as
\begin{equation} \label{eqn:r_ani}
  \beta_{\text{ani}}(r) = \frac{r^2}{r_{\text{ani}}^2+r^2}.
\end{equation}

Assuming a Hernquist profile \citep[][]{Hernquist:1990p10126} and an anisotropy radius $r_{\text{ani}}$ for the stellar orbits in the lens galaxy, the three-dimensional radial velocity dispersion $\sigma_r$ at radius $r$ from Jeans modeling is given by

\begin{multline} \label{eqn:sigma2_r}
  \sigma^2_r = \frac{4 \pi G a^{-\gamma'} \rho_0 r_0^{\gamma'}}{3 - \gamma'} \frac{r(r+a)^3}{r^2 + r_{\text{ani}}^2} \\
  \times \left( \frac{r_{\text{ani}}^2}{a^2} \frac{_2F_1 \left[ 2+\gamma',\gamma';3+\gamma'; \frac{1}{1+r/a}  \right]}{(2+\gamma')(r/a+1)^{2+\gamma'}} \right. 
  \left. + \frac{ _2F_1 \left[ 3,\gamma';1+\gamma'; -a/r  \right] }{\gamma'(r/a)^{\gamma'} } \vphantom{\frac{r_{\text{ani}}^2}{a^2} \frac{_2F_1 \left[ 2+\gamma',\gamma';3+\gamma'; \frac{1}{1+r/a}  \right]}{(2+\gamma')(r/a+1)^{2+\gamma'}}} \right),
\end{multline}

where $a$ is related to the effective radius of the lens light profile $\theta_{\text{eff}}$ by $a=0.551 \theta_{\text{eff}}$ and $_2 F_1$ is a hyper geometric function. The modeled luminosity-weighted projected velocity dispersion $\sigma_s$ is given by

\begin{equation}
  I_H(R) \sigma_s^2 = 2\int_R^{\infty} \left(1-\beta_{\text{ani}}(r)\frac{R^2}{r^2}\right) \frac{\rho_* \sigma_r^2 r dr}{\sqrt{r^2-R^2}}
\end{equation}
where $R$ is the projected radius, $\rho_*$ is the stellar density and $I_H(R)$ is the projected Hernquist distribution. The luminosity weighted LOS velocity dispersion within an aperture $\mathcal{A}$ is then (see also equation 20 in \cite{Suyu:2010p4938})
\begin{equation} \label{eqn:sigma_convolved}
  (\sigma^\text{P})^2 = \frac{\int_{\mathcal{A}}\left[I_H(R) \sigma_s^2 * \mathcal{P} \right]RdRd\theta}{\int_{\mathcal{A}}\left[I_H(R) * \mathcal{P} \right]RdRd\theta}
\end{equation}
where $* \mathcal{P}$ indicate the convolution with the seeing. In Appendix \ref{app:nummerics} we describe in detail how we compute a modeled $\sigma^\text{P}$ in a numerically stable way. This calculation assumes no rotational behaviour of the lensing galaxy. Priors on the anisotropic behaviour $\beta_{\text{ani}}(r)$ are discussed in section \ref{sec:priors}.

Equation (\ref{eqn:sigma_convolved}) can be expressed as a function of angular scales of $r_{\text{ani}}$ and $\theta_{\text{eff}}$ paired with a cosmological dependent angular diameter distance relation and an external convergence factor as
\begin{equation} \label{eqn:sigma_P}
  (\sigma^\text{P})^2 = (1-\kappa_{\text{ext}}) \cdot \frac{D_{\text{s}}}{D_{\text{ds}}} \cdot H(\gamma', \theta_{\text{E}}, \beta_{\text{ani}}(r), \theta_{\text{eff}})
\end{equation}
where $H$ is capturing all the computation of equation (\ref{eqn:sigma_convolved}) without cosmological and external convergence specifications. With this calculation, we see that any estimate of the (central) velocity dispersion is dependent on the ratio of angular diameter distance from us to the source and from the deflector to the source. This fact is important when kinematic modeling is used to infer cosmographic information. We separate in the modeling the angular and the cosmological information. The separability allows us to consistently infer cosmographic information without the need of cosmological priors in the kinematic modeling.

\subsection{Likelihood analysis} \label{sec:likelihood_analysis}
We estimate the pixel uncertainty in the image with a Gaussian background contribution $\sigma_{\text{bkgd}}$ estimated from an empty region in the image and a Poisson contribution from the model signal $d_{\text{P,i}}$ scaled by the exposure map $t_{\text{i}}$. In addition, the modeling uncertainty of the PSF of the bright point sources with amplitude $A_j$, PSF kernel $k_{ij}$ and model uncertainty coming from the star-by-star scatter $\delta_{\text{PSF}}$ is given as
\begin{equation}
  \sigma_{\text{PSF},i} = \sum_{j=1}^{N_{\text{AGN}}} A_j k_{ij} \delta_{\text{PSF},ij},
\end{equation}
at a pixel i, where $N_{\text{AGN}}$ is the number of quasar images. All together, the uncertainty for each pixel $i$ sums up in quadrature as
\begin{equation}
  \sigma_{\text{pixel},i}^2 = \sigma_{\text{bkgd}}^2 + t_{\text{i}}^{-1} d_{\text{P,i}} + \sigma_{\text{PSF,i}}^2.
\end{equation}
For the linear source surface brightness reconstruction $d_{\text{P,i}}$ is replaced by the image intensity $d_{\text{ACS,i}}$.

The likelihood of an image $\boldsymbol{d_{\text{ACS}}}$ given a model $\boldsymbol{d_{\text{P}}}$ is
\begin{equation} 
 P(\boldsymbol{d_{\text{ACS}}}| \boldsymbol{d_{\text{P}}}) = \frac{1}{Z_{\text{d}}} \exp \sum_{i=1}^{N_{\text{d}}} \left[-\frac{\left(d_{\text{ACS},i} - d_{\text{P},i}  \right)^2}{2\sigma_{\text{pixel},i}^2} \right]
\end{equation}
with $N_{\text{d}}$ being the number of pixels in the modeled image and $Z_{\text{d}}$ is the normalization
\begin{equation}
  Z_{\text{d}} = (2\pi)^{N_{\text{d}}/2} \prod_i^{N_{\text{d}}} \sigma_{\text{pixel},i}.
\end{equation}

At this stage, it is useful to separate the model into nonlinear parameters $\boldsymbol{\eta}$ and linear parameters $\boldsymbol{s}$. The likelihood of the non-linear parameters is given by 
\begin{equation} \label{eqn:likelihood_image}
  P(\boldsymbol{d_{\text{ACS}}}| \boldsymbol{\eta}) = \int d \boldsymbol{s} P(\boldsymbol{d_{\text{ACS}}}| \boldsymbol{\eta}, \boldsymbol{s}) P(\boldsymbol{s}).
\end{equation}
The integral is computed in \cite{Birrer:2015p11550} (their Equation 13) assuming flat priors in $\boldsymbol{s}$, which we adopt.

The likelihood for the time delays $\boldsymbol{\Delta t}$ is the product of the likelihoods of all relative delays of the quasar pairs $(ab)$
\begin{equation} \label{eqn:likelihood_time_delay}
  P(\boldsymbol{\Delta t}| D_{\Delta t}^{\text{model}}, \boldsymbol{\eta}) = 
  \prod_{(ab)} \left( \frac{1}{\sqrt{2\pi}\sigma_{ab}} \exp \left[-\frac{\left( \Delta t_{ab} - \Delta t_{ab}^{\text{P}}\right)^2}{2\sigma_{ab}^2}   \right] \right).
\end{equation}
The likelihood of the LOS central velocity dispersion is given by
\begin{equation} \label{eqn:likelihood_sigma_v}
  P(\sigma_v| \boldsymbol{\eta}) = \frac{1}{\sqrt{2\pi}\sigma_{\sigma}} \exp \left[-\frac{\left( \sigma_v - \sigma^{\text{P}}\right)^2}{2\sigma_{\sigma}^2}   \right].
\end{equation}

\section{The mass sheet degeneracy} \label{sec:degeneracies}
There exists many different degeneracies in strong lens modeling \citep[e.g.,][]{Saha:2000p9143, Saha:2006p9157}. In this section we focus on the MSD \cite{Falco:1985p8873} and in particular its impact on time delay cosmography as it was pointed out by \cite{Schneider:2013p8677}. As shown by \cite{Falco:1985p8873}, a remapping of a reference mass distribution $\kappa$ by 
\begin{equation} \label{eqn:mass_sheet_transform}
	\kappa_{\lambda} (\vec{\theta}) = \lambda \kappa(\vec{\theta}) + (1 - \lambda)
\end{equation}
combined with an isotropic scaling of the source plane coordinates
\begin{equation} \label{eqn:source_scaling}
 \vec{\beta} \rightarrow \lambda \vec{\beta}
\end{equation}
will result in the same dimensionless observables (image positions, image shapes and magnification ratios) regardless of the value of $\lambda$. This type of mapping is called mass-sheet-transform (MST), and shows that imaging data, no matter how good, can not break the MSD.

The additional mass term in MST (Equation \ref{eqn:mass_sheet_transform}) can be internal to the lens galaxy (affecting the lens kinematics) or due to line-of-sight structure (not affecting the lens kinematics) \citep[see e.g.,][]{Saha:2000p9143, Wucknitz:2002p10026}. The external part of the MST can be approximated by an external convergence $\kappa_{\text{ext}}$, which rescales the time delays accordingly. The external contribution also rescales the source plane. Lens modeling often only explicitly models the internal structure of the lens. The inferred source scale has to be rescaled by the external mass sheet to match the physical scale.

\subsection{Source scaling and the MSD}
An important parameter in the lens model inference is the physical source scale. Neither the lens model nor the source size are direct observables, but they share the MST in each others inference. Given a lens model, certain source sizes are preferred. The opposite is also true: Given a source size, certain lens models are preferred. This is a direct consequence of the MST (Equation \ref{eqn:mass_sheet_transform} and \ref{eqn:source_scaling}). Therefore, it is important to control the prior on the assumed source scale in the modeling.
A particular source surface brightness reconstruction method, depending on the choice of regularization, basis set, pixel grid size or parameters of the source reconstruction, will potentially favor a certain size of the reconstructed source and therefore may indirectly lead to priors on the internal mass model through the MST. As one does not know a priori the physical scales in the source galaxy, this may lead to significant biases in the inference of the lens model.

We use shapelets \citep[][]{Refregier:2003p8153} as the source surface brightness basis functions as implemented in \cite{Birrer:2015p11550}. These basis functions form a complete basis set when the order $n$ goes to infinity. When restricting the shapelet basis to a finite order $n_{\text{max}}$, the reconstruction of an image depends on the chosen scale $\beta$ of the shapelet basis function. As pointed out by \cite{Refregier:2003p8153}, for a given $n_{\text{max}}$, there is a scale $\hat\beta$ that best fits the data. From Equation \ref{eqn:source_scaling}, we see that changes in $\beta$ can be remapped into changes in the lensing potential through the linear parameter $\lambda$. Therefore, since our source reconstruction technique has an explicit scale, we have a tool to walk along the MST.

\subsection{Varying source scale in the ACS WFC1 images} \label{sec:source_scale_variation}
We have identified the source scale to have an impact on the inference of the lens model within the MST. To investigate the specific dependence of the shapelet scale in the source reconstruction in combination with lens model parameterization (Section \ref{sec:lensmodel_param}) in our analysis of RXJ1131-1231, we model the ACS WFC1 F814W and F555W images with different choices of the shapelet scale $\beta$. For the F814W image, we use the range 0.14" - 0.19" and for the F555W image the range 0.13" - 0.18". The shapelet order was held constant at $n_{\max} = 30$. To find the best fit model, we used a particle swarm optimization as used in \cite{Birrer:2015p11550} to maximize the likelihood (Equation \ref{eqn:likelihood_image}). In this section, we only use the HST images for our modeling. Time-delay and kinematic data will be added in Section \ref{sec:posterior_sampling}.

\begin{figure*}
  \centering
  \includegraphics[angle=0, width=\linewidth]{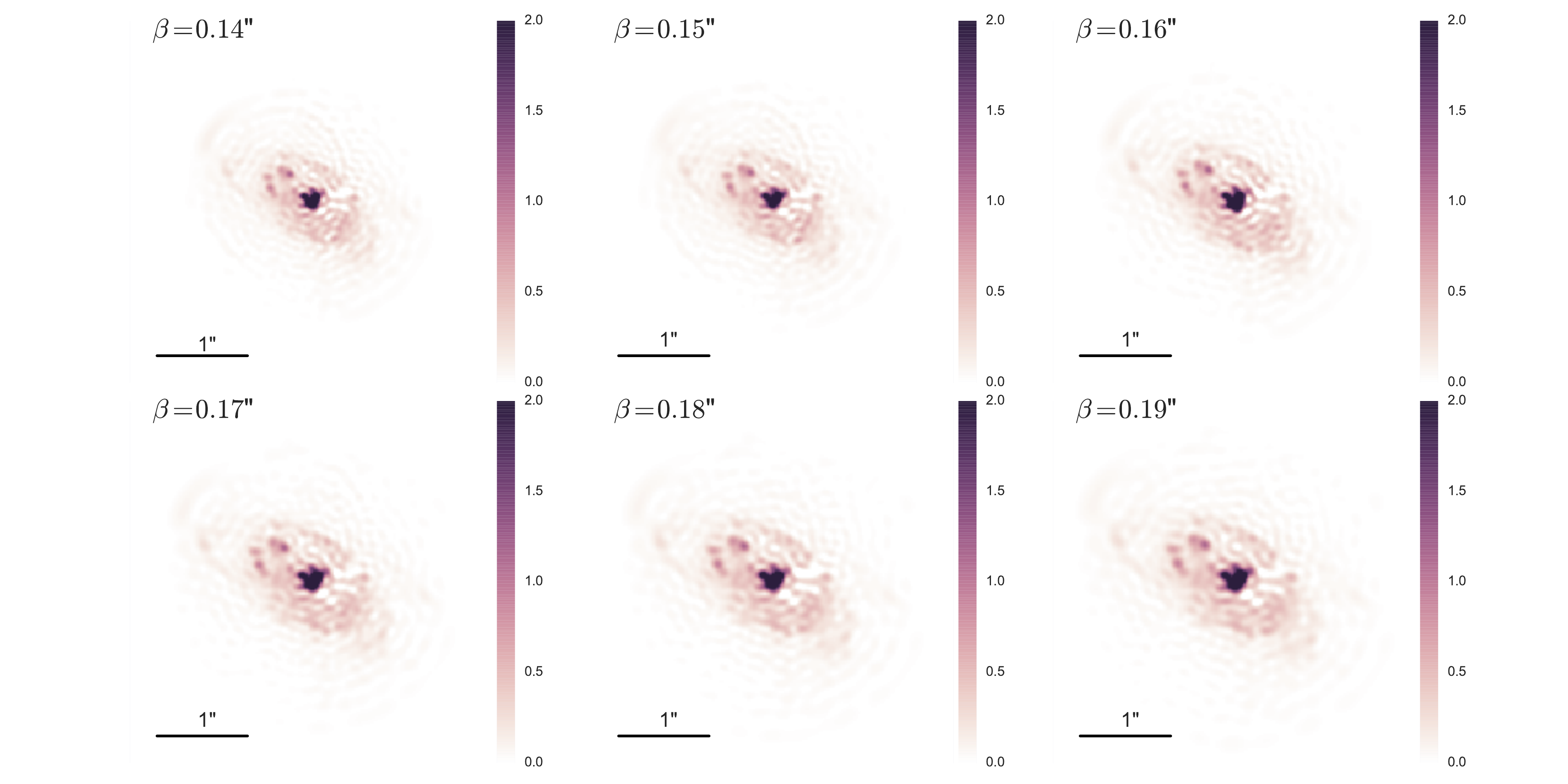}
  \caption{Reconstructed source surface brightness profiles as a function of shapelet scale $\beta$ for filter F814W. The source reconstructions of the best fit lens model configurations are shown with a given $\beta$. We see that the features become larger with larger choices of $\beta$.}
\label{fig:RXJ1131_source_beta_itter}
\end{figure*}

\begin{figure*}
  \centering
  \includegraphics[angle=0, width=50mm]{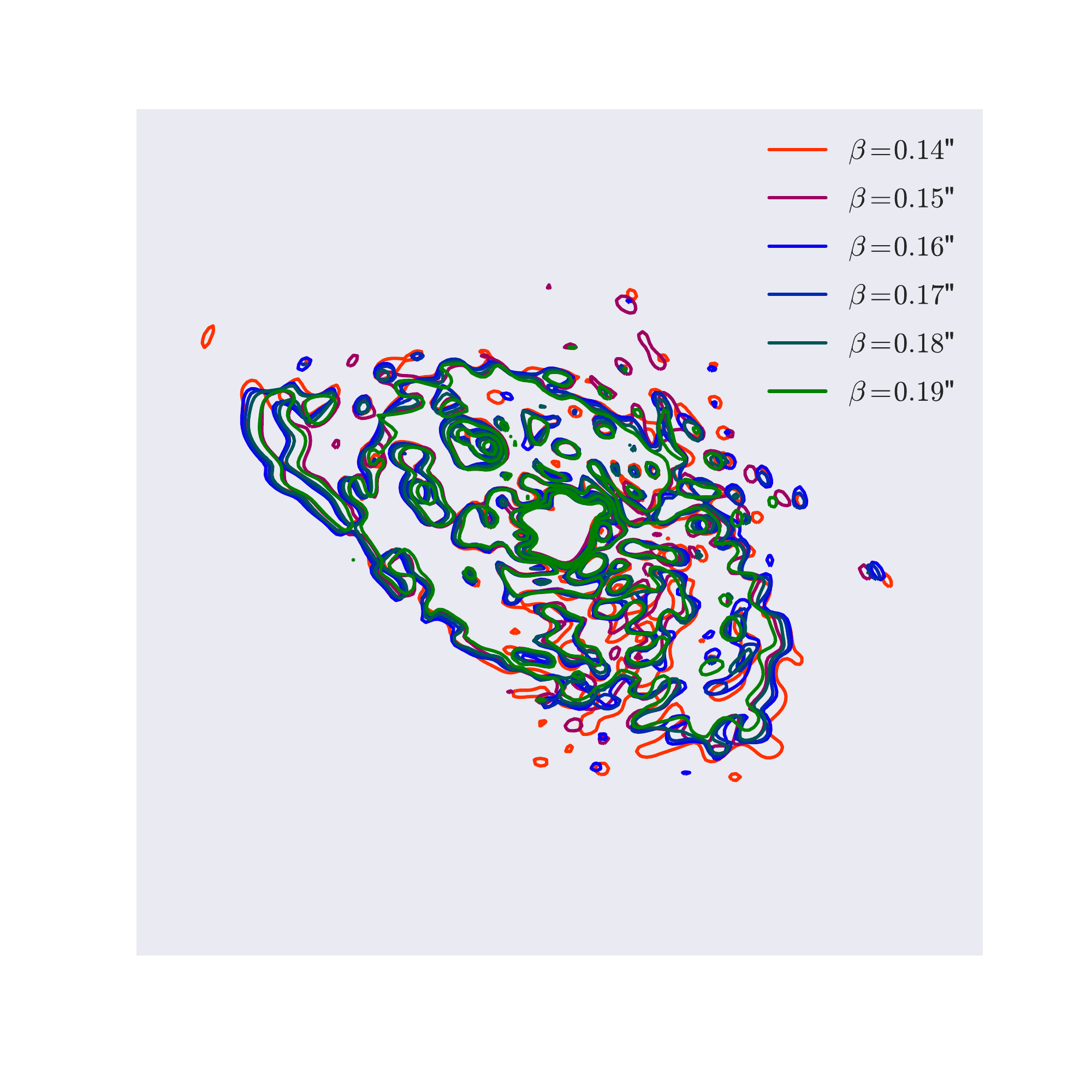}
  \includegraphics[angle=0, width=50mm]{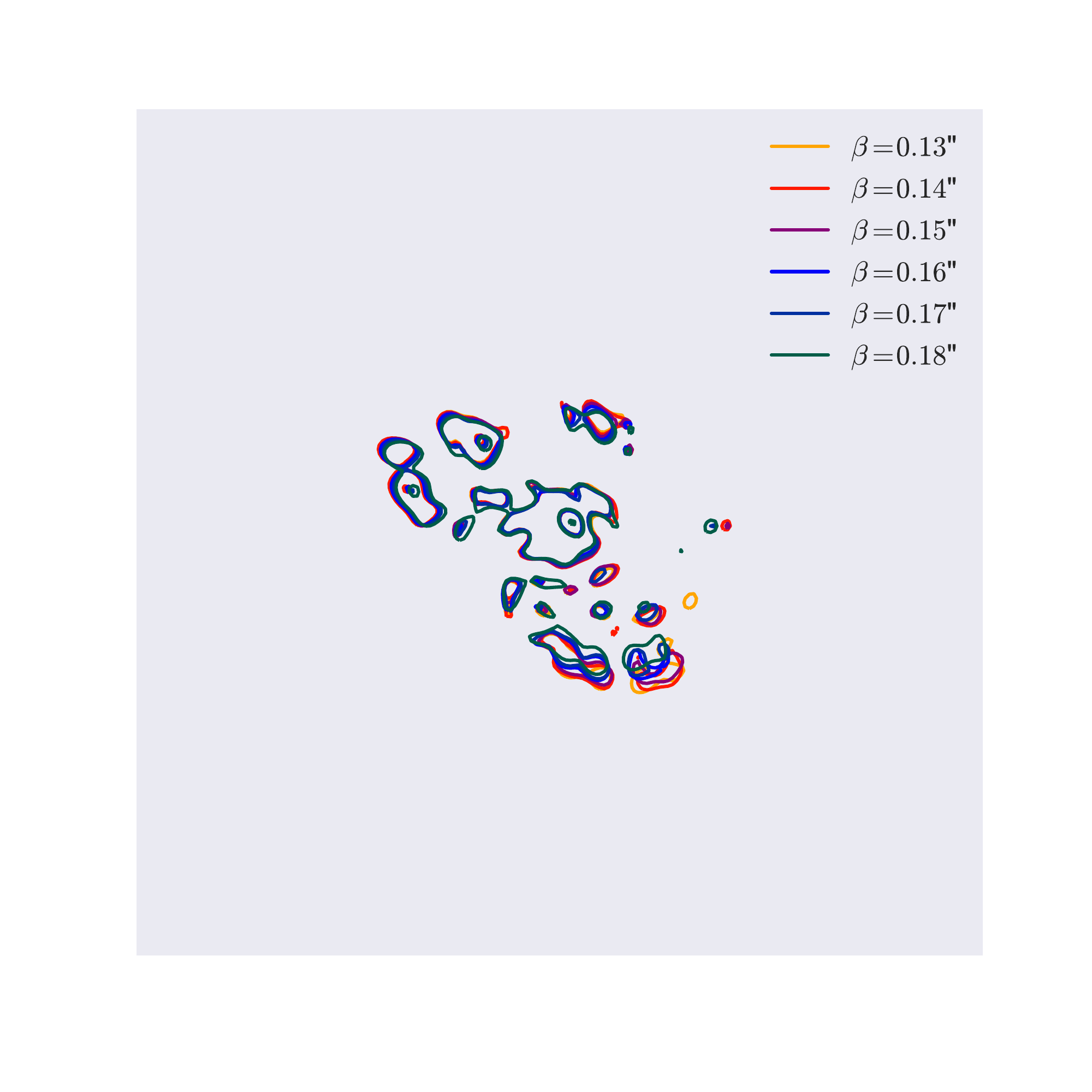}
  \includegraphics[angle=0, width=50mm]{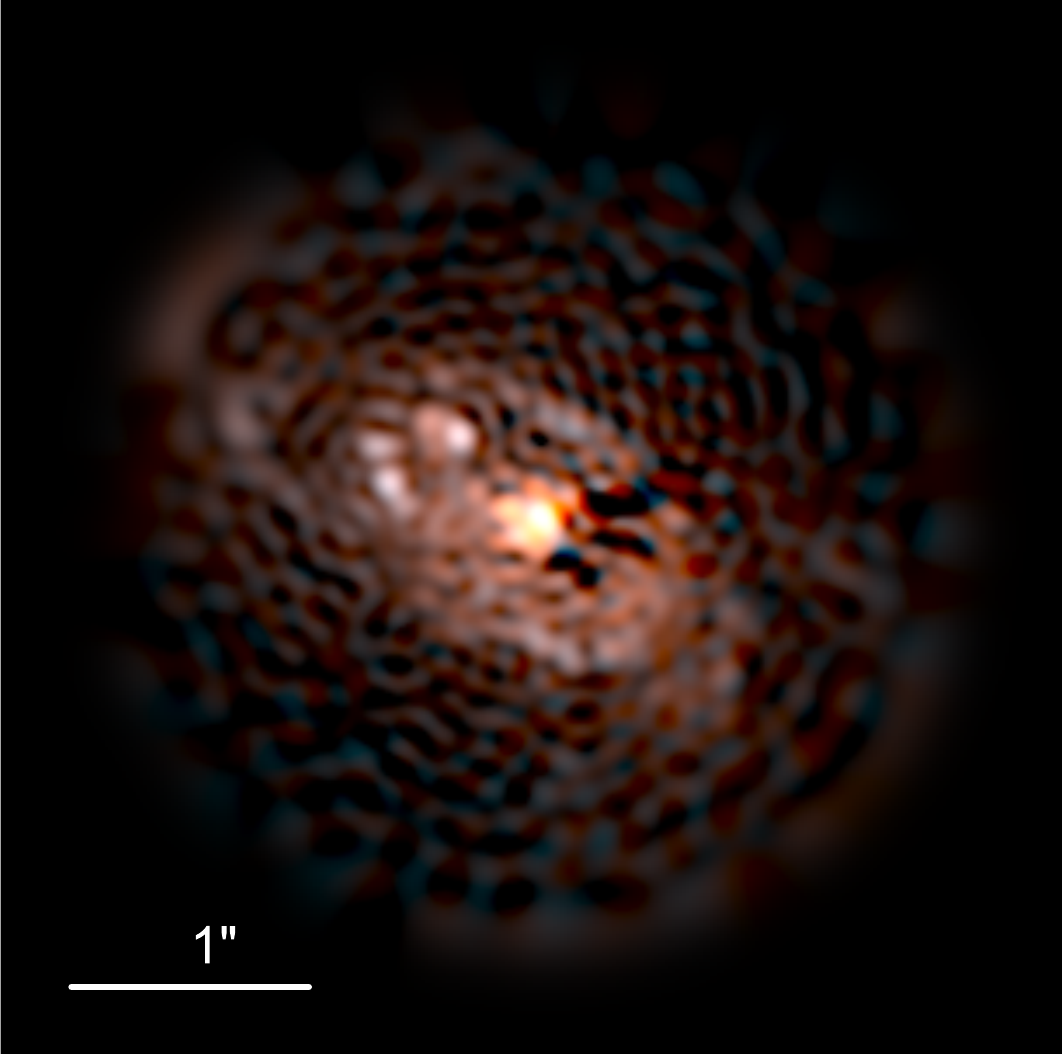}
  
  \caption{Left: Intensity contours of the reconstructed source surface profiles rescaled to fiducial value $\beta=0.2"$ for the different shapelet scales $\beta$ in filter F814W of Figure \ref{fig:RXJ1131_source_beta_itter}. The contour lines overlay well. The lens model does adopt to the choice of $\beta$ such that the source reconstruction catches the best scales. Middle: Same as left for the filter F555W. The same behavior can be seen as for F814W. Right: Color composite model of the filters F814W and F555W for a chosen joint lens model.}
  \label{fig:RXJ1131_source_contours}
\end{figure*}

Figure \ref{fig:RXJ1131_source_beta_itter} shows the source reconstruction of the best fit models of filter F814W for six different scales $\beta$. We see that the source reconstructions are very similar but scaled by the relative factors of the chosen shapelet scale. More explicitly, we overlay in Figure \ref{fig:RXJ1131_source_contours} the intensity contours of the different source reconstructions rescaled by $\beta$. We also show the reconstructions for the F555W image, which shows the same behavior. On the right of Figure \ref{fig:RXJ1131_source_contours} we over-plot a joint source reconstruction of the two bands in a fake color image. In Appendix \ref{app:residuals}, we present the corresponding normalized residuals for this analysis of the F814W reconstruction.

The difference in the likelihood value for different scales $\beta$ from the imaging data exceeds the 10-$\sigma$ level between each modeled scale $\beta$. This reflect the fact that the chosen lens model parameterization (see Section \ref{sec:lensmodel_param}) does not allow for the full freedom needed for a perfect transform according to the MST (Equation \ref{eqn:mass_sheet_transform}). The source scale $\beta$ can not be fixed to an arbitrary value and caution on any scale dependent source reconstruction description is needed. When assigning a prior on $\beta$ and infer this parameter together with all the lens model parameters from the image reconstruction, we are able to very precisely determine the corresponding source scale and the parameters of the given functional form of the lens model.

\subsection{Relaxing on the lens model assumption} \label{sec:relax_lens_model}
As pointed out by \cite{Schneider:2013p8677}, there can also be an internal component to the MST. Namely when the lens model can not reproduce the underlining internal mass distribution. The assumption of a power-law lens model formally sets the internal part of the MST. The parameters will fit preferentially those models, whose shape, modulo an artificial MST, are the most similar to the underlying mass distribution. The only effect visible in the modeling of the imaging data is on the source scale. The inferred source scale will be different from the one of the true lens model. Any assumed mass distribution which can not be rescaled according to Equation (\ref{eqn:mass_sheet_transform}) can thus potentially lead to biased inferences, in particular on the slope of the mass profile. This also can result in significant biases in the inferred lensing potential and lens kinematics. In particular, it was stated by \cite{Schneider:2013p8677} that the assumption of a power-law lens model can potentially lead to a significant bias in the inference of the time delay distance.

Three approaches to handle the concerns of \cite{Schneider:2013p8677} in performing cosmographic estimates are:
\begin{enumerate}
  \item One assumes that the true lens model can be described within the functional form of the chosen parameterization. This is the approach done by \cite{Suyu:2013p4952}. In this case we end up with the potentially biased inference discussed in \cite{Schneider:2013p8677}, a situation we want to avoid as good as possible.

  \item One choses a more flexible lens model than a single power-law mass profile. This approach was followed in \cite{Suyu:2014p8316} in response to \cite{Schneider:2013p8677}. Different profile parameterizations may lead to different preferred source scales. It is not guaranteed that a more sophisticated lens model parameterization infers an unbiased result in the cosmographic inference.

  \item Perform simplifications and approximations that lead to greater robustness against known degeneracies. For instance accommodating MST through careful handling of the source size inference.

\end{enumerate}
In this work we chose the third option mentioned above. This option requires the least assumptions on the lens model and a prior is placed on the source size, rather through the functional form of the lens model. In Appendix \ref{app:renormalization_bayes} we specifically state the process in a Bayesian inference way to make clear our steps and approximations and show that a renormalization of the imaging likelihood for different imposed source scales $\beta$ is needed to explore the impact of plausible internal MST on the cosmological inference.

\subsection{Adding lens kinematics} \label{sec:breaking_msd}
Additional constraints on the lens model can come from kinematic data at a different scale than the Einstein ring. This becomes of particular importance when weakening the constraining power of the lens model, as described in Section \ref{sec:relax_lens_model}. Lens models with different source scales predict different lens kinematics. The prediction depends on the stellar velocity anisotropy $\beta_{\text{ani}}$ which can not be known from the existing data and the external convergence $\kappa_{\text{ext}}$ which has to be inferred separately.

As long as the relative likelihood of additional kinematic data (Equation \ref{eqn:likelihood_sigma_v}) can not compete with the relative likelihood of the different shapelet scales $\beta$ (on the 10-$\sigma$ level between the chosen source scales, see Section \ref{sec:source_scale_variation}), the combined likelihood will be dominated by the lens model assumption. Only when re-normalizing the likelihood of the imaging data for different scales $\beta$, the kinematic data can have a significant impact in the determination of the lens profile and in particular the lens potential for time-delay cosmography.

\section{Combined likelihood analysis} \label{sec:posterior_sampling}
In this section, we discuss how we combine the different data sets and their likelihoods. We showed in the previous section that biases can emerge from choices in the lens and source modeling. These aspects have to be taken into account when the data sets are combined.

\subsection{Combining imaging and time delay data}
In a first step, we do a joint analysis of the independent measurements of the time delay and imaging data. The combined likelihood is
\begin{equation} \label{eqn:likelihood_image_delay}
  P(\boldsymbol{d_{\text{ACS}}}, \boldsymbol{\Delta t}| \boldsymbol{\eta}, D_{\Delta t}^{\text{model}}) = P(\boldsymbol{d_{\text{ACS}}}| \boldsymbol{\eta}) P(\boldsymbol{\Delta t}| D_{\Delta t}^{\text{model}}, \boldsymbol{\eta})
\end{equation}
with the independent likelihoods of Equation (\ref{eqn:likelihood_image}) and (\ref{eqn:likelihood_time_delay}). We do not yet combine the kinematic data at the likelihood level. We sample all the lens model parameters and the time delay distance $D_{\Delta t}^{\text{model}}$. We keep the lens light parameters fixed at the final position of the particle swarm process in the MCMC process to achieve a more efficient sampling of the relevant parameters. We included the full flexibility of the lens light parameters on a subset of the MCMC chains and come to the conclusion that the additional covariance of the lens light model on the cosmographic analysis is very minor, i.e. the impact on the uncertainty on $H_0$ is below 0.1\%.

From Bayes theorem, the likelihood of the parameters given the data is (modulo a normalization):
\begin{equation}
  P(\boldsymbol{\eta}, D_{\Delta t}^{\text{model}}| \boldsymbol{d_{\text{ACS}}}, \boldsymbol{\Delta t}) 
  \propto P(\boldsymbol{d_{\text{ACS}}}, \boldsymbol{\Delta t}| \boldsymbol{\eta}, D_{\Delta t}^{\text{model}}) P(\boldsymbol{\eta}) P(D_{\Delta t}^{\text{model}}).
\end{equation}
We apply flat priors on the parameters $\gamma' \in [1,2.8]$, $\theta_{E} \in [0.1", 10"]$, $q \in [0.5,1]$, $\theta_{E,\text{clump}} \in [0",1"]$, $\gamma_{\text{ext}} \in [0,0.3]$ and $D_{\Delta t}^{\text{model}} \in [0, 10'000]$ Mpc.

At this stage, we want to emphasize that there are 3 data points in the time delay measurement compared to several thousands of high signal-to-noise pixels in the imaging comparison. In principle, the provided time delay measurement can not only determine $D_{\Delta t}^{\text{model}}$, which is independent of the imaging data but also can partially constrain the lens model. In practice, any even minor bias introduced in the image modeling can out-weigh the constraining power of the two additional time delay measurements.

\begin{figure*}
  \centering
  \includegraphics[angle=0, width=150mm]{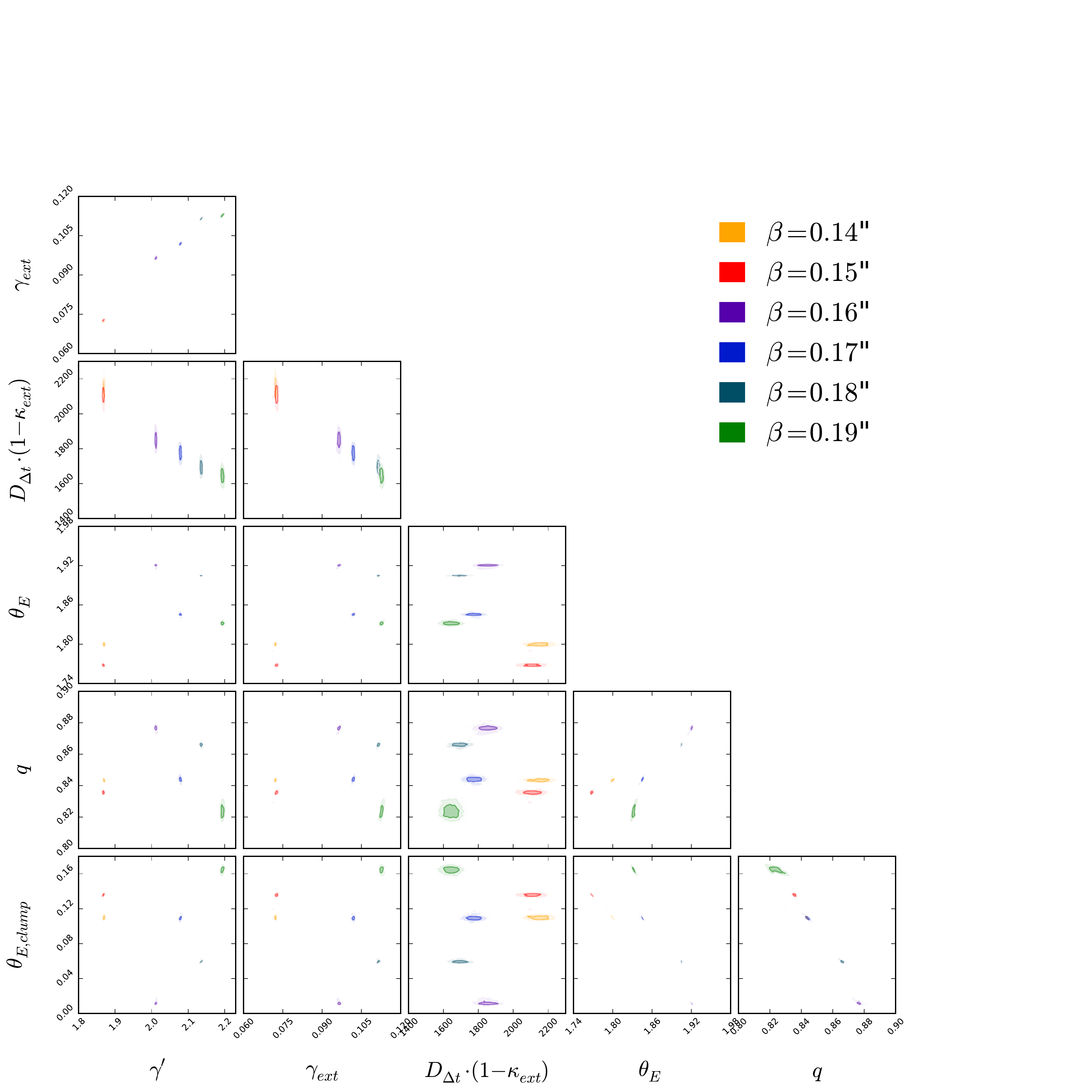}
  \caption{Posterior distribution (1-2-3 sigma contours) of lens model parameters and time delay distance of the combined analysis of imaging data of F814W and time delay measurements. Different colors correspond to different choices of the shapelet scale $\beta$. The posterior samples for different $\beta$ values mutually disagree in almost all parameters presented.}
\label{fig:RXJ1131_constraints}
\end{figure*}

In the following, we present the results of the analysis of filter F814W. The results of the equivalent analysis of filter F555W can be found in Appendix \ref{app:F555W}. To sample the posterior distribution of the parameter space we use \texttt{CosmoHammer} \citep[][]{Akeret:2013p8317}. We fix the shapelet scale $\beta$ at [0.14", 0.15", 0.16", 0.17", 0.18", 0.19"] and do a separate inference of the parameters for each choice of $\beta$. Figure \ref{fig:RXJ1131_constraints} shows the posterior distribution of some of the parameters for the different choices of $\beta$. The inferred parameter constraints for different $\beta$ values do not overlap. We see that $\gamma_{\text{ext}}$ is very narrowly determined for a given shapelet scale $\beta$ but varies from $0.07$ up to $0.11$ depending on the position in the degeneracy plane. We want to stress that the external convergence $\kappa_{\text{ext}}$ estimated by \cite{Suyu:2013p4952} is based on an external shear prior of $\gamma_{\text{ext}} = 0.089 \pm 0.006$.

\subsection{Constraints from kinematic data} \label{sec:kin_constraints}
To investigate the potential constraining power of the velocity dispersion data, we are interested in how distinguishable different positions within the MST are in terms of their predicted central velocity dispersions. To do so, we fix the cosmology and the external convergence $\kappa_{\text{ext}}$ to fiducial values. This allows us to evaluate the predicted LOS central velocity dispersion $\sigma^{\text{P}}$ (Equation \ref{eqn:sigma_P}) for all the posterior samples of Figure \ref{fig:RXJ1131_constraints}. We assume a random realization of $r_{\text{ani}}$ with a flat prior in the range [0.5,5]$\theta_{\text{eff}}$ for all the posterior positions.

\begin{figure}
  \centering
  \includegraphics[angle=0, width=90mm]{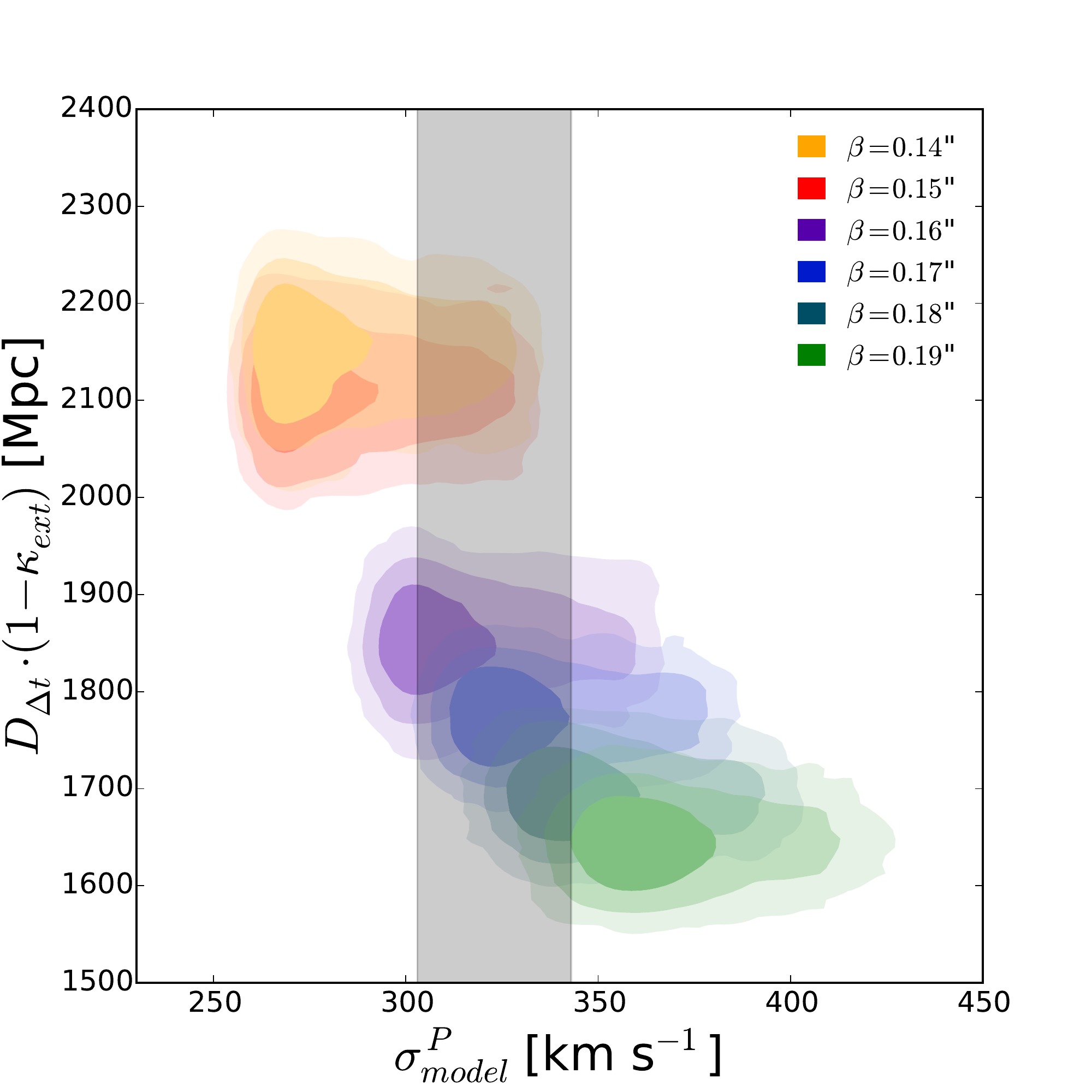}
  \caption{Estimated LOS central velocity dispersions $\sigma^{\text{P}}$ vs. time delay distances $D_{\Delta t}$ of the sample of lens models from Figure \ref{fig:RXJ1131_constraints} (in the same colors) for a kinematic anisotropy prior of $P_{[0.5,5]}(r_{\text{ani}})$. The 1-2-3 sigma contours are shown. The external convergence $\kappa_{\text{ext}}$ was explicitly set to zero and the cosmology has been fixed to the Planck mean values in this particular plot. The gray band reflects the 1-$\sigma$ uncertainty range of the LOS velocity dispersion estimates from the data. This shows that velocity dispersion estimates add important information on the lens model constraints.}
\label{fig:RXJ1131_D_t_sigma}
\end{figure}

In Figure \ref{fig:RXJ1131_D_t_sigma} we illustrate the predicted $\sigma^{\text{P}}$ samples vs the predicted time delay distance $D_{\Delta t}$. We see that the samples can not be fully distinguished with the current velocity dispersion measurement and the assumed anisotropy prior. The relative distance in the predicted velocity dispersion $\sigma^{\text{P}}$ between the different samples are all within 4$\sigma$ (model given data).

There are three factors which affects the distinction of the source scales by kinematic data. (1) The uncertainty in the spectroscopic measurement, analysis and modeling of $\pm 20$km s$^{-1}$ which is about 6\%. This is visually the most obvious contribution in Figure \ref{fig:RXJ1131_D_t_sigma}, marked by the gray band. The mean values of the predicted samples of the different source scales differ by about one sigma of this estimated uncertainty. (2) The anisotropic uncertainty in the lens galaxy kinematics. This is the main driver of the spread in the predictions of the velocity dispersion within each source scale sample. This scatter has a relative spread of 10\% given $P_{[0.5,5]}(r_{\text{ani}})$. (3) The predicted velocity dispersion depend highly on the observational conditions and configuration. The PSF and the slit size of the spectrograph results in a convolution and averaging over a wide range of radial scales. The predicted velocity dispersion for different concentrations of the mass in the lens galaxy (i.e. power-law slope $\gamma'$) differ the most in the very center of the lens. At the Einstein radius itself, the different lens models predict basically the same kinematics. With the PSF of 0.7" and a slit size of 0.81"$\times$ 0.7", power-law mass profiles with slopes in the range $\gamma' \in [1.8,2.2]$ differ by about 100 km s$^{-1}$ in their predicted velocity dispersion $\sigma^{\text{P}}$. A smaller slit and seeing conditions of FWHM 0.1" can double this relative difference and therefore could improve the constraining power of the kinematic data significantly.

The combined effect of non-perfect data and non-perfect modeling of the kinematic data with prior $P_{[0.5,5]}(r_{\text{ani}})$ can be translated in a relative error in the time delay distance $D_{\Delta t}$ of about \kinematicsError from Figure \ref{fig:RXJ1131_D_t_sigma}. Only kinematic data of the lens galaxy and its analysis can reduce this error budget.

In Section \ref{sec:source_scale_variation} we showed that the individual image likelihoods of the different $\beta$ samples differ by more than 10$\sigma$. Before including the velocity dispersion measurement in our cosmographic analysis, we re-normalize the image likelihood such that it is independent of the source scale $\beta$ (see Section \ref{sec:relax_lens_model}). This re-normalization is done by taking the same number of MCMC posterior samples from the different source scales $\beta$ when doing further inferences with the lens model parameters.

\subsection{Source scale and kinematic anisotropy priors} \label{sec:priors}
The combination and inference coming from the different data sets relies on priors on the source scale of the background galaxy and on the anisotropic behaviour of the stellar kinematics in the lens galaxy. In particular, the inference of the Hubble constant $H_0$ is related to the inference of the angular diameter distance $D_{\Delta t}$ as
\begin{equation}
  H_0 \propto D_{\Delta t}^{-1}.
\end{equation}
In Figure \ref{fig:RXJ1131_D_t_sigma}, we see a significant dependence between the size of the source galaxy ($\propto \beta$) and $D_{\Delta t}$. Furthermore the interpretation of the kinematic data is also dependent on the anisotropic behaviour of the lens galaxy.

Choices of the priors on the source size $P(\beta)$ and aniosotropic kinematic $P(\beta_{\text{ani}}(r))$ must be chosen with care based on information gained from other work as these priors potentially have a significant impact on the infered parameter posterior (i.e. $H_0$). In the following, we discuss two different priors in the kinematic anisotropy and the source scale. \footnote{Comments from the authors about confirmation bias: The analysis of the mentioned priors on the cosmological inference has been made after posting a first version of this paper on arXiv.}

\subsubsection{Source size prior $P(\beta)$} \label{sec:source_scale_prior}
A simple form of the source size prior which does not impose any specific form of knowledge about $\beta$ is a uniform prior in the range $[0", 10"]$. We refer to this prior as $P_{\text{flat}}(\beta)$. This prior ignores any knowledge about the population of galaxies. The model parameter $\beta$ is directly related to the brightness $L$ of the source as
\begin{equation}
  \beta^2 \propto L.
\end{equation}
The number density of galaxies as a function of luminosity is a well measured quantity (luminosity function, LF) and its faint end slope for the blue galaxy population can be well described with a single power-law slope as
\begin{equation}
  \frac{dn}{dL} \propto L^{\alpha_{\text{LF}}}
\end{equation}
with $\alpha_{\text{LF}} = -1.30$ \cite{Faber:2007p11257}. In this form, the expected source size can be stated as
\begin{equation}
  P_{\text{LF}}(\beta) = \frac{dn}{d\beta} = \frac{dn}{dL} \frac{dL}{d\beta} \propto \beta^{2\alpha_{\text{LF}}+1}.
\end{equation}
This prior is weakly dependent on $\beta$ such that smaller source sizes are prefered. We chose $P_{\text{flat}}(\beta)$ as our default prior and explore the impact with $P_{\text{LF}}(\beta)$ in section \ref{sec:prior_dependent_inference}.

\subsubsection{Anisotropic kinematic prior $P(\beta_{\text{ani}}(r))$}
Studies of early type (lens) galaxies have been made by e.g. \cite{Koopmans:2009p11307, Barnabe:2011p11334} which reveal similar properties compared to local early type galaxies.
We consider two priors which cover the same range in the mean anisotropic behaviour and their predicted velocity dispersion $\sigma^{\text{P}}$. (1) The prior used in Figure \ref{fig:RXJ1131_D_t_sigma} is flat in $r_{\text{ani}}$ (equation \ref{eqn:r_ani}) in the range [0.5,5]$\theta_{\text{eff}}$. This prior should cover the expected scale where the transition between isotropic and radial velocity dispersion should occur in an uniform way and is exactly the same prior used in \cite{Suyu:2013p4952}. We refer to this prior as $P_{[0.5,5]}(r_{\text{ani}})$.

(2) We model a global contribution of the anisotropic behaviour in the form
\begin{equation} \label{eqn:b_anisotropy}
  \beta_{\text{ani}} = 1 -  \frac{\bar{\sigma_z^2}}{\bar{\sigma_r^2}} \equiv 1 - \frac{1}{b}
\end{equation}
in the range $[1, 1.5]$. This reflects the same range in allowed $\sigma^{\text{P}}$ values for a given mass model. We refer to this prior as $P_{[1,1.5]}(b)$. $b = 1$ indicates a isotropic velocity dispersion and $b = 1.5$, for which the velocity dispersion ellipsoid is very elongated along the radial direction with $\beta_{\text{ani}} = 0.33$, corresponds to $r_{\text{ani}}= 0.5 \theta_{\text{eff}}$ with the same mean anisotropy within the aperture. This is the same functional form of the prior as used in \cite{Barnabe:2012p11353} to analyze a spiral lens galaxy althought with less range into a pure radial dispersion.

\section{Cosmological inference} \label{sec:cosmological_inference}
In this section, we study the cosmological constraints from strong lensing using data from images, time delays, central velocity dispersion of the lensing galaxy and independent external convergence estimates. We first show that the data can be used to constrain the angular diameter relation. Based on the constraints on the angular diameter distances, we then introduce the likelihood that allows us to infer the parameters within the flat $\Lambda$CDM cosmological model.

\subsection{Angular diameter distance posteriors}
We can combine the posterior samples of Figure \ref{fig:RXJ1131_constraints} with the independent velocity dispersion measurement to calculate the angular diameter distance relations $D_{\text{d}}$ and $D_{\text{s}}/D_{\text{ds}}$ (Equation \ref{eqn:sigma_P} and \ref{eqn:D_t_definition}) as
\begin{equation}
  \frac{D_{\text{s}}}{D_{\text{ds}}} = \frac{(\sigma^\text{P})^2}{(1-\kappa_{\text{ext}})}\frac{1}{H(\gamma', \theta_{\text{E}}, \beta_{\text{ani}}(r), \theta_{\text{eff}})}
\end{equation}
and
\begin{equation}
  D_{\text{d}} = \frac{D_{\Delta t}^{\text{model}}}{(1 + z_d)(1 - \kappa_{\text{ext}})} \frac{D_{\text{ds}}}{D_{\text{s}}}. 
\end{equation}
To take into account the errors in $\sigma_v$, $\kappa_{\text{ext}}$ and $r_{\text{ani}}$, we importance sample the posteriors from the independent measurements ($\sigma_v$ and $\kappa_{\text{ext}}$) and for $r_{\text{ani}}$ we uniformly sample in the range [0.5,5] times $\theta_{\text{eff}}$ \citep[see e.g.][for similar use]{Lewis:2002p10133, Suyu:2010p4938, Suyu:2013p4952}.

\begin{figure*}
  \centering
  \includegraphics[angle=0, width=75mm]{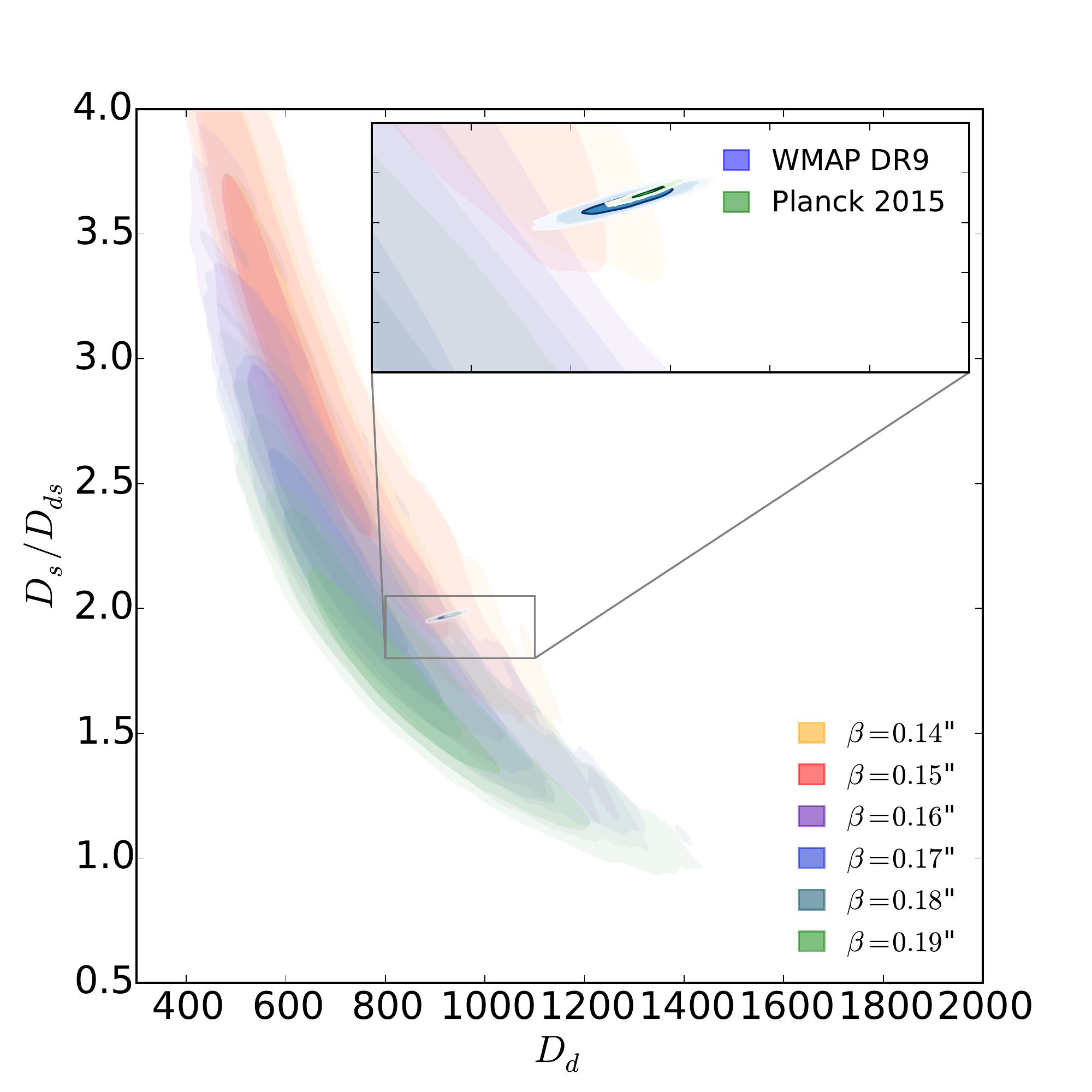}
  \includegraphics[angle=0, width=75mm]{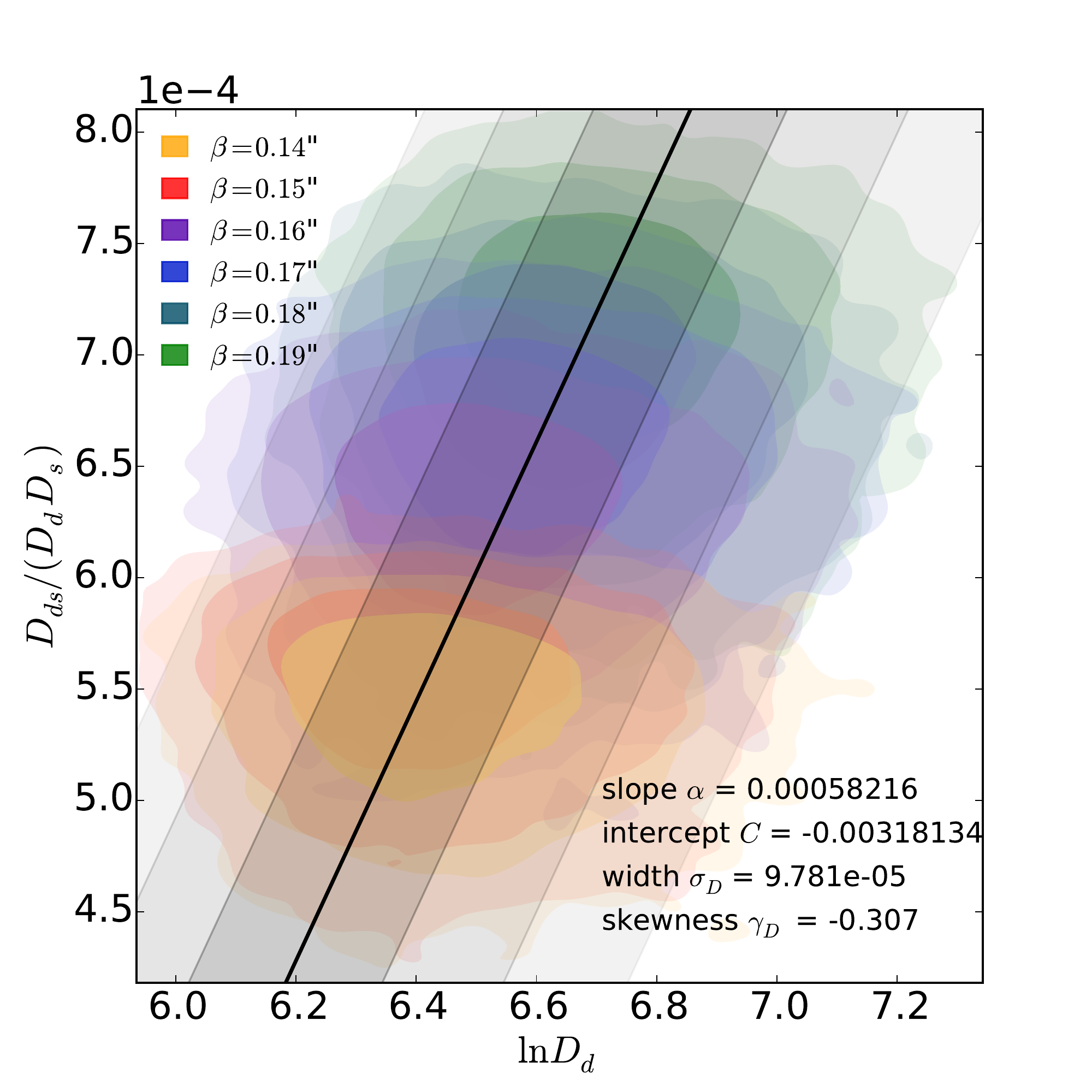}

  \caption{The constraints of the angular diameter distance relation for discrete positions in the MSD plane for filter F814W (same analysis for filter F555W is shown in Figure \ref{fig:RXJ1131_F555W_Dd_Ds_Dsd} in the appendix). The chosen priors in the source scale and the kinematic anisotropy of the lensing galaxy are $P_{\text{flat}}(\beta)$and $P_{[0.5,5]}(r_{\text{ani}})$. Different colors indicate different imposed source scales. On the left panel: $D_d$ vs $D_s/D_{ds}$. Also over-plotted are the posteriors of the WMAP DR9 and Planck 2015 $\Lambda$CDM posteriors mapped in the same angular diameter distance relation. On the right panel: Re-mapping of the angular diameter relations into a $\ln D_{\text{d}}$ vs $D_{\text{ds}}/(D_{\text{d}} D_{\text{s}})$ plane. The linear fit is indicated by the thick black line and the (1,2,3)-$\sigma$ upper and lower limits of the projected distribution are plotted in different gray scale. The parameters of the fit are indicated in the figure.}
  \label{fig:RXJ1131_Dd_Ds_Dsd}
\end{figure*}

The $D_{\text{d}}$ vs $D_{\text{s}}/D_{\text{ds}}$ plane as shown in Figure \ref{fig:RXJ1131_Dd_Ds_Dsd} inherits the cosmological information of this analysis coming from the combined data and consistently translates the uniform prior in the source scale into the cosmological inference. This plane covers a wide range but the constrained region is more narrow. \cite{Jee:2015p10857} did a very similar analysis in term of folding in the velocity dispersion measurement. In our case, we get a degeneracy in the two-dimensional plane coming from the MST whereas \cite{Jee:2015p10857} and the forecasting of \cite{Jee:2015p10858} assume independence in the two quantities. We over-plot the posterior samples of WMAP DR9 \citep[][]{Hinshaw:2013p7705} and Planck15 \citep[][]{PlanckCollaboration:2015p9875} converted to the angular diameter distances of the lens system. We find that at least the posterior samples of one chosen source scale parameter $\beta$ is consistent within 2$\sigma$ with the CMB experiment posteriors in a flat $\Lambda$CDM cosmology for the low redshift angular diameter distance relations. Without the renormalization of the imaging likelihood (see Section \ref{sec:relax_lens_model}), this statement can not be made.

\subsection{An analytic likelihood for cosmology}
So far, we have discretized the degeneracy plane by uniformly sample $\beta$ in steps of 0.01". Effectively this means that while all the other parameters are sampled through standard MCMC methods, the $\beta$ direction is sampled on a grid. This separation is needed to allow us to do the re-normalization of the likelihood as described in Section \ref{sec:kin_constraints}. Sampling the $\beta$-grid finely is computationally expensive. In the following, we show how we can analytically describe the posterior distribution and fill the gaps in $\beta$ without additional sampling.

To do so, we first map the $D_{\text{d}}$ vs $D_{\text{s}}/D_{\text{ds}}$ plane of Figure \ref{fig:RXJ1131_Dd_Ds_Dsd} (left panel) into a $\ln D_{\text{d}}$ vs $D_{\text{ds}}/(D_{\text{d}} D_{\text{s}})$ plane (right panel). We see a linear relations between the posterior samples in a monotonic and equally spaced increasing fashion as a function of $\beta$. We fit with linear regression the function
\begin{equation} \label{eqn:fitting}
  \frac{D_{\text{ds}}}{D_{\text{d}} D_{\text{s}}} = \alpha \ln(D_{\text{d}}) + C
\end{equation}
with $\alpha$ being the slope and $C$ being the intercept. The legend of Figure \ref{fig:RXJ1131_Dd_Ds_Dsd} (right panel) shows the best fit values, which we discuss in more detail later. The linear fit is a good description of the combined samples of different source scalings. The same is shown for the filter F555W analysis in Appendix \ref{app:F555W}. The spread of the distribution orthogonal to the linear relation is not well fit by a Gaussian distribution, but we find a skewed normal distribution provides a good description.

The one-dimensional likelihood $P(\boldsymbol{d_{\text{RXJ}}}, \boldsymbol{\pi})$ of the strong lens system data $\boldsymbol{d_{\text{RXJ}}}$ given a cosmological model $\boldsymbol{\pi}$ is given by the one-dimensional probability density of the samples relative to the fitted line:
\begin{equation} \label{eqn:skew_likelihood}
  P(\boldsymbol{d_{\text{RXJ}}}, \boldsymbol{\pi}) = \phi_{\gamma} \left( x=\frac{D_{\text{ds}}}{D_{\text{d}} D_{\text{s}}}, \mu = \alpha \ln (D_{\text{d}}) + C, \sigma_D, \gamma_D \right),
\end{equation}
where $\sigma_D$ is the standard deviation, $\gamma_D$ the skewness and $\phi_{\gamma}$ is the re-parameterized skewed normal distribution function described in Appendix \ref{app:skew_normal_distribution}. How the different source scale priors on $\beta$ fold in the likelihood is described in Appendix \ref{app:source_size_prior} and equation \ref{eqn:source_size_prior}. In this section, we apply a flat prior in $\beta$, $P_{\text{flat}}(\beta)$, and a flat prior in $r_{\text{ani}}$, $P_{[0.5,5]}(r_{\text{ani}})$, (see section \ref{sec:priors}). The inferences for the other combinations of the choices of priors are presented in Section \ref{sec:prior_dependent_inference}.

For the analysis of the HST band F814W we fit the values $C=\interceptRed $, $\alpha=\alphaRed $, $\sigma_D=\sigmaRed$ and $\gamma_D = \skewnessRed$. For band F555W the fits result in $C=\interceptBlue $, $\alpha=\alphaBlue $, $\sigma_D=\sigmaBlue$ and $\gamma_D = \skewnessBlue$. Fitting the combined samples of the band F814W and F555W leads to $C=\interceptBoth $, $\alpha=\alphaBoth $, $\sigma_D=\sigmaBoth$ and $\gamma_D = \skewnessBoth$. The units of these parameters are given in respect with the angular diameter distances in Mpc.

The simple form of the likelihood enables a fast and consistent combination of different strong lensing systems also in combination with other cosmological probes.

\subsection{Cosmological parameter constraints}
\begin{figure*}
  \centering
  \includegraphics[angle=0, width=160mm]{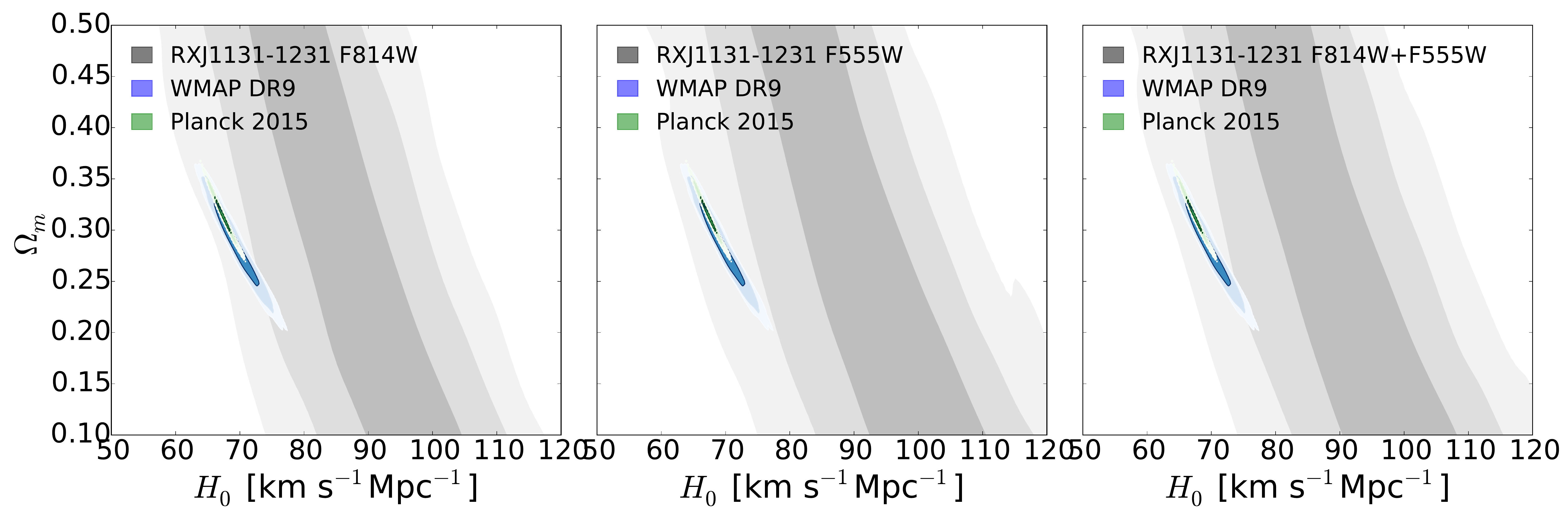}
  \caption{Posterior sampling of the cosmological parameters for the filters F814W (left), F555W (middle) and combined with equal weight of the likelihoods of the two images (right). The posterior distribution of WMAP DR9 and Planck 2015 are over-plotted. The chosen priors in the source scale and the kinematic anisotropy of the lensing galaxy are $P_{\text{flat}}(\beta)$and $P_{[0.5,5]}(r_{\text{ani}})$.}
\label{fig:H0_omega_m_constraints}
\end{figure*}

The constraints on the angular diameter distance relations can be turned into constraints on the cosmological parameters of the background evolution. In the following we assume a flat $\Lambda$CDM cosmology. The homogeneous expansion can be described in terms of the matter density $\Omega_m$ and the Hubble constant $H_0$. We use the likelihood of Equation (\ref{eqn:skew_likelihood}) with the values of $\alpha$, $C$, $\sigma_D$ and $\gamma_D$ from the analysis of F814W and F555W separately. First, we sample the parameters $\Omega_m$ and $H_0$ simultaneously with uniform priors of $\Omega_m \in [0,1]$ and $H_0 \in [0, 200]$. Figure \ref{fig:H0_omega_m_constraints} shows the posterior distributions for the filter F814W (left panel) and F555W (middle panel) for the priors ($P_{\text{flat}}(\beta)$, $P_{[0.5,5]}(r_{\text{ani}})$) separately. The degeneracy in $\Omega_m$ is strong but $H_0$ can be determined fairly well. A good approximation of the degeneracy shown in the $H_0$-$\Omega_m$-plane can be described by

\begin{equation} \label{eqn:H0_omega_m_degeneracy}
  H_0 = H_0^*\left[1 + \frac{1}{2}(\Omega_m-\Omega_m^*)\right]^{-1} \pm \sigma_{H_0^*} \left(\frac{H_0}{H_0^*}\right)
\end{equation}
where $H_0^*$ is the value for $H_0$ at fixed $\Omega_m^*$ and $\sigma_{H_0^*}$ is the marginalized error at fixed $\Omega_m^*$. This form allows us to more directly compare with other results from the literature.

For a fixed value of $\Omega_m = 0.3$, we infer a Hubble constant of $H_0=\HFilterRed$ \hubbleUnit for the F814W and $H_0=\HFilterBlue$ \hubbleUnit for the F555W analysis.

\begin{figure}
  \centering
  \includegraphics[angle=0, width=90mm]{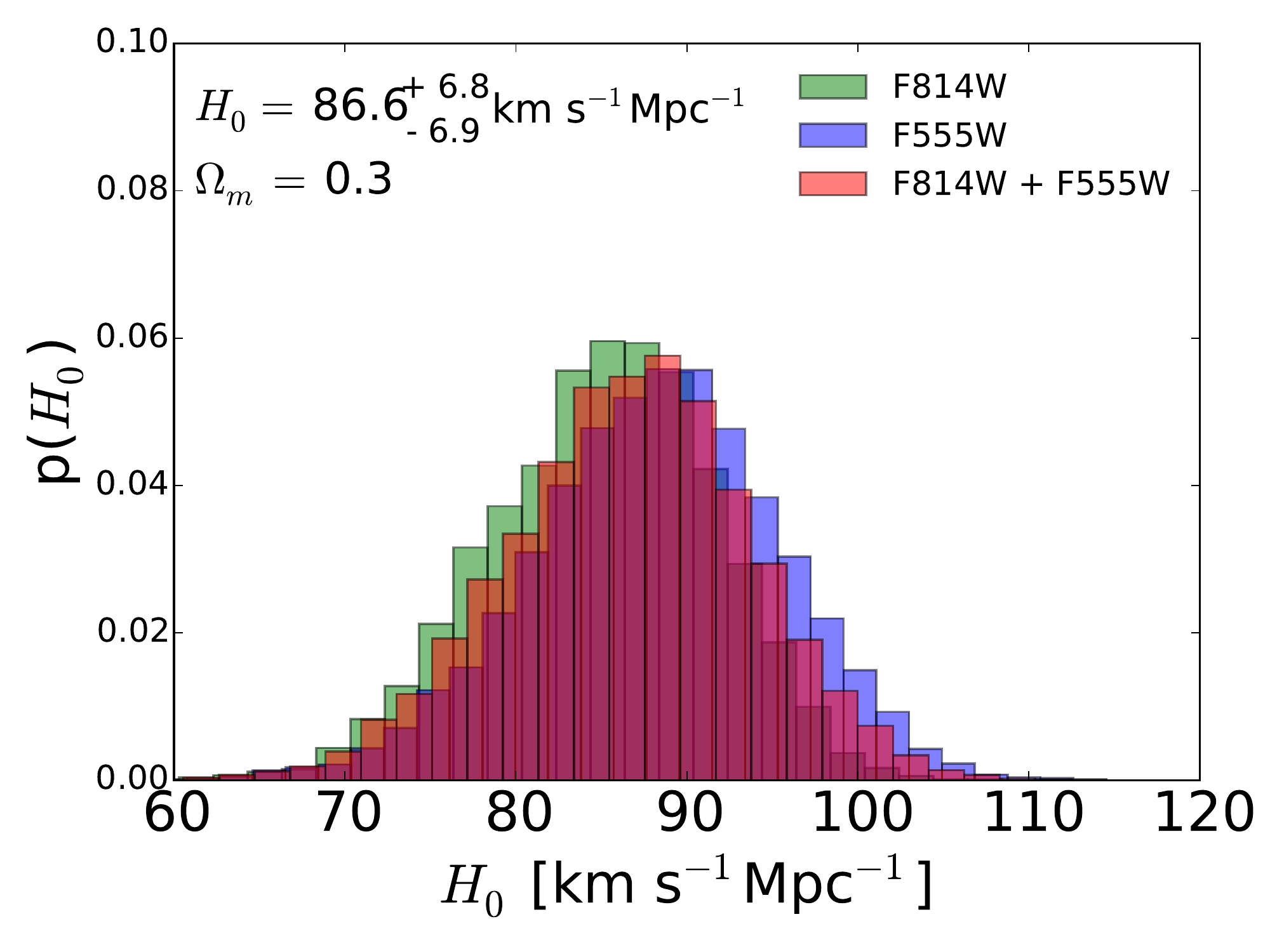}
  \caption{Posterior distribution for the value of $H_0$ for a fixed $\Omega_m=0.3$ for filter F814W (green), F555W (blue) and the combined samples (red). The chosen priors in the source scale and the kinematic anisotropy of the lensing galaxy are $P_{\text{flat}}(\beta)$and $P_{[0.5,5]}(r_{\text{ani}})$.}
\label{fig:H0_constraints}
\end{figure}

From the analysis of each filter separately, we get an uncertainty coming from the imaging data only to be below 1\% in the resulting inference of $H_0$. Given the fact that our estimates for the two filters F814W and F555W is about 4.0\% different while using exactly the same analysis and the same time-delay and kinematic data for all other parameters involved, we conclude that the imaging data inference is partially driven by unknown systematics in the modeling and the data. To marginalize out potential systematics in the analysis, we combine the two analyses on the angular diameter posterior level. The two-dimensional posteriors are shown in the right panel of Figure \ref{fig:H0_omega_m_constraints}. In this way, we get a Hubble constant of $H_0=\HFilterBoth$ \hubbleUnit. The full posteriors for both samples are shown in Figure \ref{fig:H0_constraints}.

\subsection{Prior dependence} \label{sec:prior_dependent_inference}
In this section we investigate the dependence of the cosmological inference from the choice of priors of the source scale $\beta$ and the anisotropic kinematics of the lensing galaxy $\beta_{\text{ani}}$. In Section \ref{sec:priors} we stated for each parameter two different priors, each of them being quoted to be uninformative and probing the same range in the physics. In table \ref{tab:prior_dependence} the likelihood parameters and the resulting $H_0$ inference for fixed $\Omega_m = 0.3$ in flat $\Lambda$CDM are stated. We see a strong prior dependence on the posterior distribution which can result in a mean shift in $H_0$ of more than 10 \hubbleUnit. The source scale prior $P(\beta)$ can result in a weak mean shift of about 1-2 \hubbleUnit without a change in the uncertainty. This means that the information content in the imprinted priors are roughly the same and the systematic uncertainty is subdominant to the quoted total uncertainty. The situation changes for the kinematic prior $P(\beta_{\text{ani}})$. The flat prior approach for the two different parameterizations shifts the mean infered value of $H_0$ by more than $1\sigma$. The precision is also affected: The prior $P_{[0.5,5]}(r_{\text{ani}})$ results in a significantly higher precision inference than $P_{[1,1.5]}(b)$. This implies that $P_{[0.5,5]}(r_{\text{ani}})$ inherits more information for the specific task of measuring $H_0$ than $P_{[1,1.5]}(b)$. If this prior is not representative of the distribution of early type galaxies, the inference with $P_{[0.5,5]}(r_{\text{ani}})$ can be significantly biased compared with $P_{[1,1.5]}(b)$.

\begin{table}[t]
  \centering 

  \begin{tabular}{llrrrrrrr}
  \hline \hline
  $P(\beta_{\text{ani}})$ & $P(\beta)$ & $C$ & $\alpha$ & $\sigma_D$ & $\gamma_D$ & $C_{\beta}$ & $\alpha_{\beta}$ & $H_0$\tablefootnote{For fixed $\Omega_m = 0.3$ in flat $\Lambda$CDM.}\\
  \hline
  $P_{[0.5,5]}(r_{\text{ani}})$  & $P_{\text{flat}}(\beta)$ & $\interceptBoth$ & $\alphaBoth$ & $\sigmaBoth$ & $\skewnessBoth$ & - & - & $\HFilterBoth$\\
  $P_{[0.5,5]}(r_{\text{ani}})$ & $P_{\text{LF}}(\beta)$  & $\interceptBoth$ & $\alphaBoth$ & $\sigmaBoth$ & $\skewnessBoth$ & -0.0012 & 263.8 & $\HBothFlatSource$\\
  $P_{[1,1.5]}(b)$ & $P_{\text{flat}}(\beta)$  & $\interceptBothAni$ & $\alphaBothAni$ & $\sigmaBothAni$ & $\skewnessBothAni$ & - & - & $\HBothAniFlatBeta$ \\
  $P_{[1,1.5]}(b)$ & $P_{\text{LF}}(\beta)$ & $\interceptBothAni$ & $\alphaBothAni$ & $\sigmaBothAni$ & $\skewnessBothAni$ & -0.0014 & 264.2 & $\HBothAniFlatSource$ \\

  \hline \hline
  \end{tabular}
  \caption{Likelihood and posteriors for different choices of priors. The $H_0$ inference is for fixed $\Omega_m = 0.3$. $P_{[0.5,5]}(r_{\text{ani}})$ indicates a flat prior in $r_{\text{ani}}$ in the range $[0.5,5]\theta_{\text{eff}}$ in the parameterization of equation \ref{eqn:r_ani} and $P_{[1,1.5]}(b)$ indicates a flat prior in $b$ of equation \ref{eqn:b_anisotropy} of the anisotropic behavior of the lens galaxy. $P_{\text{flat}}(\beta)$ reflects a flat prior in the source scale and $P_{\text{LF}}(\beta)$ reflects a prior of the galaxy luminosity function (see section \ref{sec:source_scale_prior}). The parameters describe the likelihood function stated in equation \ref{eqn:skew_likelihood} and \ref{eqn:source_size_prior}.}
  \label{tab:prior_dependence}
\end{table}

\section{Joint uncertainties and comparison with other work} \label{sec:comparison}
In this Section we analyze the impact of the different data sets on the cosmological inference and we compare our method and results with the literature.

\subsection{Uncertainties from the different data sets}
We assign uncertainty estimates on the inference of $H_0$ coming from the independent data sets, namely the time delays, the HST ACS images, the line-of-sight analysis of wide field data and the spectra of the lens galaxy for the kinematic estimate \footnote{In this analysis we ignore the dependence of the line-of-sight analysis on the shear term from the ACS image reconstruction.}. We do so by forecasting a perfect modeling result for all data sets except the one in question. We then proceed in exactly the same way as presented in Section \ref{sec:cosmological_inference}. This leads to an inference of the cosmological parameters only affected by the uncertainties coming from one single data set. We perform this analysis with the default priors $P_{[0.5,5]}(r_{\text{ani}})$ and $P_{\text{flat}}(\beta)$.

In Table \ref{tab:error_budget} the estimated uncertainties from the different data sets are summarized and the 1-$\sigma$ uncertainties on $H_0$ for fixed $\Omega_m$ is stated. The Gaussian approximation of all these errors leads to a total uncertainty of \totalErrorGaussian on $H_0$. The estimate of the uncertainty coming from the full sampling results in \totalErrorFull. This analysis does not include further potential systematics and does not question the priors chosen.

Our approach on the error analysis is different than the one chosen by \cite{Suyu:2013p4952}. We do not quote an error on the lens model itself, as this inference is dependent on different data sets. We quote an error on the lens model modulo a MST for the image reconstruction and separately an error on the kinematic estimate, which potentially can fully break the degeneracy.

We clearly see that the dominant contribution in the final uncertainty can be related to the kinematic data and its modeling. As discussed in Section \ref{sec:kin_constraints}, high resolution spectroscopy can provide data which can better constrain different positions in the MST and therefore significantly reduce the uncertainty on the angular diameter distance relation. The second most dominant uncertainty come from the line-of-sight contribution.

\subsection{Comparison with other work}
Cosmographic inference has been published by \cite{Suyu:2013p4952} with the same lens model parameterization and by \cite{Suyu:2014p8316} in combination with a composite (dark matter and baryonic matter separated) lens model, in response to the work of \cite{Schneider:2013p8677}. The values and uncertainties on the Hubble constant are $H_0 = \HubbleSuyua$ \hubbleUnit for a value of $\Omega_m = 0.27$ in \cite{Suyu:2013p4952}, a 5.5\% error, and $H_0 = \HubbleSuyub$ \hubbleUnit, a 5.75\% error, with $\Omega_m = 0.27$ for a flat $\Lambda$CDM universe.

One difference between the work of \citep[][]{Suyu:2013p4952, Suyu:2014p8316} and the one presented in this work arise from the explicit treatment of the MSD and related degeneracies in our work and its link to the source surface brightness reconstruction method. This allows us to overcome (at least partially) systematics from the source reconstruction method and the mass profile assumption. On the other hand, this weakens the constraining power of the image reconstruction. This explains our larger uncertainties compared to \citep[][]{Suyu:2013p4952, Suyu:2014p8316}. Furthermore, their stated values on $H_0$ are $\Omega_m$-independent in the flat scenario while our values do depend on $\Omega_m$ (see our Figure \ref{fig:H0_omega_m_constraints} vs. Figure 8 in \cite{Suyu:2013p4952}). This comes from the different description of the cosmological likelihood. The likelihood in \cite{Suyu:2013p4952} is described fully in terms of the time-delay distance $D_{\Delta t}$ where else our likelihood has an additional dependence on $D_d$. In that sense, their stated $H_0$ value is independent of $\Omega_m$ but ours requires a prior on $\Omega_m$.

A second difference is that we work in a 2D-plane of angular diameter distance relations (Figure \ref{fig:RXJ1131_Dd_Ds_Dsd}) without the need of cosmological priors to define our angular diameter distance likelihood. This results in a different shape of the posterior distribution in the $\Omega_m$-$H_0$ plane (Figure \ref{fig:H0_omega_m_constraints}) and the inferred projected $H_0$ posteriors have a strong $\Omega_m$ dependence.

The best comparison with the work of \citep[][]{Suyu:2013p4952, Suyu:2014p8316} should be done when comparing the inference with the same kinematic prior $P_{[0.5,5]}(r_{\text{ani}})$ (first or second row in Table \ref{tab:prior_dependence}). We want to stress that we use explicit priors on the source scale. The cosmological inference is dependent on this prior as the constraining power of the kinematic data is weak. Therefore a shift of about 1$\sigma$ in our stated uncertainty on the inference of $H_0$ is not surprising. 

Comparing our results with the CMB experiments, we get a 2.5 $\sigma$ shift for $P_{[0.5,5]}(r_{\text{ani}})$ and a 1$\sigma$ shift for $P_{[1,1.5]}(b)$ in the $\Lambda$CDM parameter inference. We conclude that the angular diameter distance at last scattering and the inferred angular diameter distance relation at lower redshift from this analysis are consistent with a flat $\Lambda$CDM cosmology. Our analysis depends on uninformative priors on the kinematics of the lens galaxy  $\beta_{\text{ani}}$ and the source reconstruction scale $\beta$. Further systematics can potentially also occur and are not included in this analysis.

\begin{table}[t]
  \centering 
  \caption{Error budget on $H_0$ for a fixed $\Omega_m$.}

  \begin{tabular}{lr}
  \hline \hline
  Description & Uncertainty \\
  \hline
  Time delays  & \timeDelayError \\
  HST ACS image reconstruction & \lensError \\
  Line-of-sight contribution & \losError \\
  Lens kinematics \tablefootnote{The quoted uncertainty includes the uncertainty in the unisotropy radius $r_{\text{ani}}$ with a prior of $[0.5, 5]\theta_{\text{eff}}$.} & \kinematicsError \\
  \hline
  Total (Gaussian) & \totalErrorGaussian\\
  Total (full sampling \tablefootnote{The uncertainty in the full sampling is given as half of the 68\% confidence interval divided by the mean posterior value.}) & \totalErrorFull \\
  \hline \hline

  \end{tabular}
  \label{tab:error_budget}
\end{table}

\section{Conclusions} \label{sec:summary}
In this work we applied the newly developed source reconstruction technique of \cite{Birrer:2015p11550} to the strong lens system RXJ1131-1231 to extract cosmographic information. We showed how different source reconstruction scales probe different regimes in the MST even when the lens model is not fully transformable through the MST.

This work is built on the modeling and the data of \cite{Suyu:2013p4952} and the systematics analysis of \cite{Schneider:2013p8677}. We incorporate a re-normalization of the imaging likelihood such that we have explicit priors on the source scale before combining with the kinematic data.

We introduced a cosmographic inference analysis which enables us to combine imaging, time-delay and kinematic data without relying on any cosmological priors. We came up with a likelihood function only based on the angular diameter distance relations, which can be described in analytic terms.

We find that the choice of priors on lens model parameters and source size are subdominant for the statistical errors for $H_0$ measurements of this systems. The choice of prior for the source is sub-dominant at present (2\% uncertainty on $H_0$) but may be relevant for future studies. More importantly, we find that the priors on the kinematic anisotropy of the lens galaxy have a significant impact on our cosmological inference. When incorporating all the above modeling uncertainties, we find  $H_0 = \HFilterBoth$ \hubbleUnit (for $\Omega_m = 0.3$), when using kinematic priors similar to other studies. When we use a different kinematic prior motivated by Barnab\`e et al. (2012) \cite{Barnabe:2012p11353} but covering the same anisotropic range, we find $H_0 = \HBothAniFlatSource$ \hubbleUnit. This means that the choice of kinematic modeling and priors have a significant impact on cosmographic inferences. Further systematics in the data and modeling can also occur. The way forward is either to get better velocity dispersion measures which would down weight the impact of the priors or to construct physically motivated priors for the velocity dispersion model.

This inference analysis was achieved with a single strong lens system in two imaging bands. Combining the information of multiple systems with comparable data can add vital constraints about the late time expansion history of the universe, also in terms of extensions of the standard cosmological model.

\acknowledgments
We thank Sherry Suyu and the co-authors of \cite{Suyu:2013p4952} and \cite{Suyu:2014p8316} for useful comments and discussions. We thank the Referee for useful comments on the manuscript that helped us improving the text. We acknowledge the import, partial use or inspiration of the following python packages: CosmoHammer \citep[][]{Akeret:2013p8319}, FASTELL \citep[][]{Barkana:1998p5324}, numpy \footnote{www.numpy.org}, scipy \footnote{www.scipy.org}, astropy \footnote{www.astropy.org}, triangle \footnote{https://github.com/dfm/triangle.py}. This work has been supported by the Swiss National Science Foundation (grant 200021\_149442/1 and 200021\_143906/1).

\bibliography{papers_bibtex}

\appendix

\section{Numerical computation of the luminosity-weighted LOS velocity dispersion} \label{app:nummerics}

The computation of the luminosity-weighted LOS velocity dispersion within an aperture under certain seeing conditions $\sigma^\text{P}$ (Equation \ref{eqn:sigma_convolved}) involves numerically challenging projection integrals and convolutions. In this section, we describe our approach to achieve a numerically stable and fast computation with a Monte-Carlo ray-tracing approach, similarly used by e.g. \cite{Berge:2013p5329} to render convolved Galaxy light profiles. This method is based on drawing positions representing the total light distribution of the galaxy.

For the light in the galaxy, we take a Hernquist profile \cite{Hernquist:1990p10126}
\begin{equation} \label{eqn:Hernquist}
  I(r) = \frac{I_0 a}{2\pi r(r+a)^3}
\end{equation}
where $I_0$ is the total flux and $a$ relaxed to the effective radius of the galaxy by $a=0.551 \theta_{\text{eff}}$. The radial distribution function of flux is then
\begin{equation}
  P(r)dr = \frac{2r}{(r+a)^3} dr.
\end{equation}
The cumulative distribution function is
\begin{equation}
  P(<r) = \int_0^r \frac{2r'}{(r'+a)^3} dr' = \frac{r^2}{(a+r^2)}.
\end{equation}
A sample of $P(r)$ can then be drawn from the distribution
\begin{equation} \label{eqn:P_r}
  P(r) = \frac{a \sqrt{\mathcal{U}}\left( \sqrt{\mathcal{U}} + 1 \right)}{1 - \mathcal{U}},
\end{equation}
where $\mathcal{U}$ is the uniform distribution in $[0,1]$.

In the following, we describe the steps starting from a representative sample of the flux in the galaxy to get to the estimate of the aperture averaged velocity dispersion:

\begin{enumerate}
  \item Draw a representative sample of radii $r_i$ drawn from the three-dimensional light distribution of the Hernquist profile (Equation \ref{eqn:P_r}).
  \item Project the radius $r_i$ on a random two-dimensional plane and compute its projected radius $R_i$ and the projected coordinates $(x_i, y_i)$. This sample represents the projected light profile of the galaxy.
  \item Displace the two-dimensional coordinates $(x_i, y_i)$ with a random realization according to the seeing distribution to $(x_i',y_i')$. We assume the PSF is a two-dimensional Gaussian distribution. This sample represents the convolved, projected two-dimensional light distribution of the galaxy.
  \item Select samples, whose displaced position is on the aperture $(x_i',y_i') \in \mathcal{A}$. This selects a sample representative for the luminosity and radial weighting within the aperture.
  \item Evaluate $\sigma_s^2(r_i, R_i)$, the projected (but unweighted) velocity dispersion for the remaining samples.
  \item Take the sample average of the velocity dispersion $\langle \sigma_s^2(r_i, R_i) \rangle$. This average (once converged) corresponds to $(\sigma^{\text{P}})^2$ with the assumption of a Gaussian velocity dispersion.
\end{enumerate}
About 100 samples evaluated in the aperture gives already an accuracy in $\sigma^{\text{P}}$ of about 1\%. For this paper, the computation is done with 1000 samples.

\section{Residual maps} \label{app:residuals}
\begin{figure*}
  \centering
  \includegraphics[angle=0, width=\linewidth]{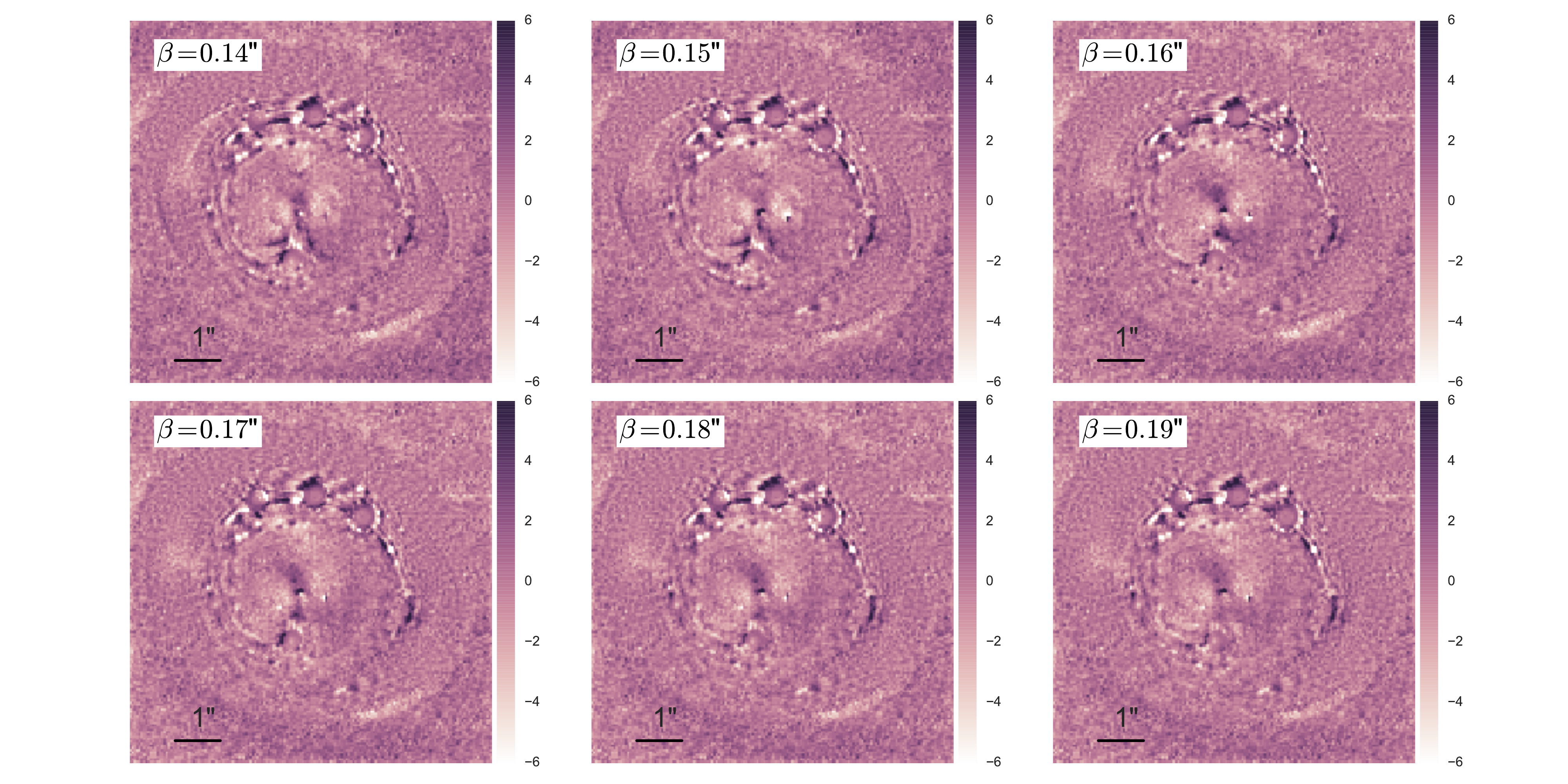}
  \caption{The normalized residual maps for the best fit reconstruction for the different choices of the shapelet scale $\beta$ for the F814W image. The residuals differ significantly for the different choices of $\beta$. From the imaging data only, a scale $\beta=0.19"$ is favored over a scale $\beta=0.14"$ by more than 30 $\sigma$. This statement is entirely lens model dependent.}
\label{fig:RXJ1131_source_beta_itter_residuals}
\end{figure*}

In Figure \ref{fig:RXJ1131_source_beta_itter_residuals} the normalized residuals corresponding to the source models with different source scales $\beta$ in Section \ref{sec:source_scale_variation} are shown. The residual maps differ significantly between the best fit values of the different shapelet scales $\beta$. This reflects the fact that extended structure in the Einstein ring can give constraints on the local slope of the mass profile and the given mass model can not adopt equally well to different source scales as it is can not be rescaled according to the mass-sheet transform. The inferred lens models can be understood as the best fit power-law profiles at different positions within the MST.

\section{Analysis on WFC1 F555W}\label{app:F555W}
\begin{figure*}
  \centering
  \includegraphics[angle=0, width=75mm]{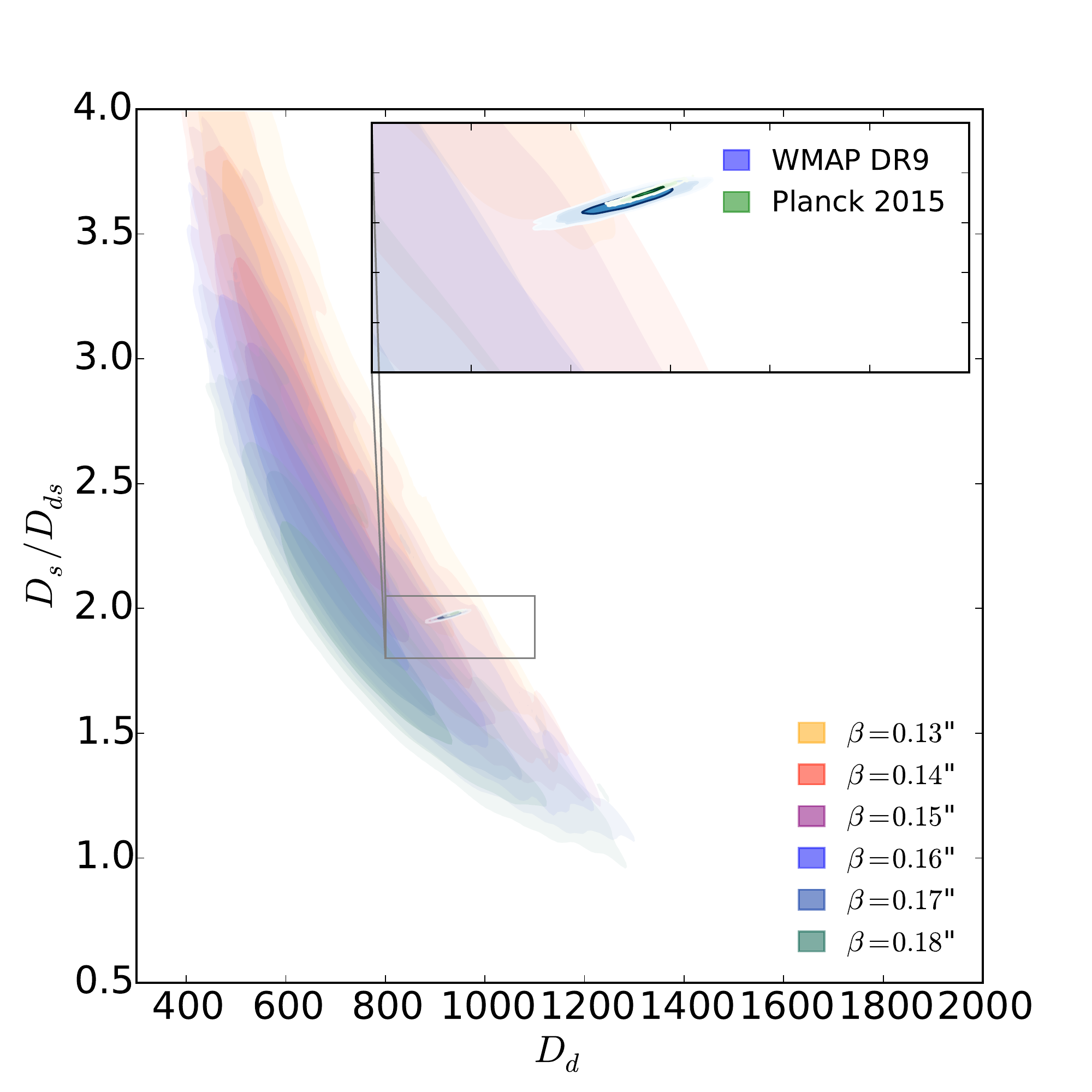}
  \includegraphics[angle=0, width=75mm]{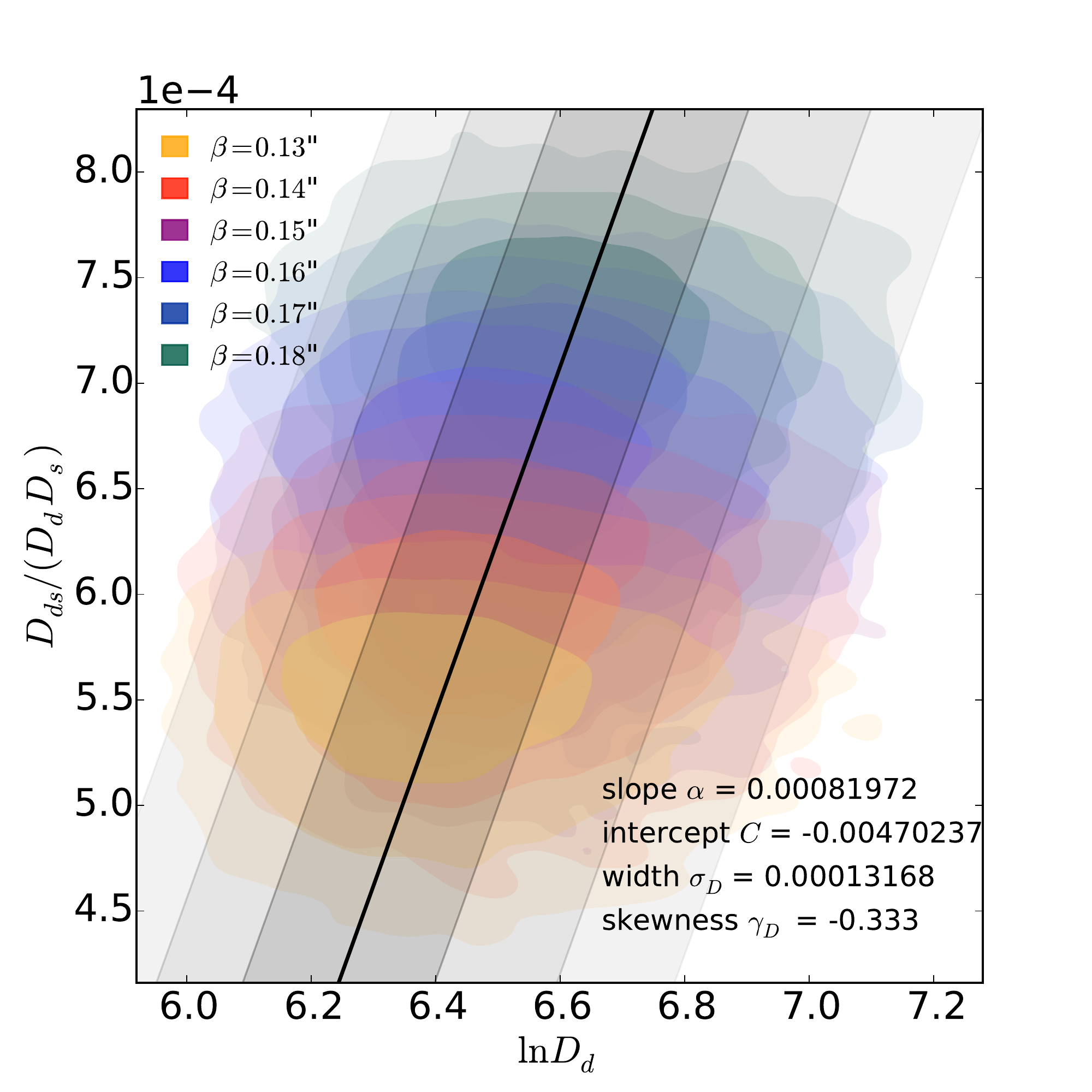}
  \caption{The constraints of the angular diameter distance relation for discrete positions in the MSD plane for filter F555W (same as Figure \ref{fig:RXJ1131_Dd_Ds_Dsd} for filter F814W). Different colors indicate different imposed source scales. On the left panel: $D_d$ vs $D_s/D_{ds}$. Also over-plotted are the posteriors of the WMAP DR9 and Planck 2015 $\Lambda$CDM posteriors mapped in the same angular diameter distance relation. On the right panel: Re-mapping of the angular diameter relations into a $\ln D_{\text{d}}$ vs $D_{\text{ds}}/(D_{\text{d}} D_{\text{s}})$ plane. The linear fit is indicated by the thick black line and the (1,2,3)-$\sigma$ upper and lower limits of the projected distribution are plotted in gray scale. The parameters of the fit are indicated in the figure.}
\label{fig:RXJ1131_F555W_Dd_Ds_Dsd}
\end{figure*}

\begin{figure*}
  \centering
  \includegraphics[angle=0, width=150mm]{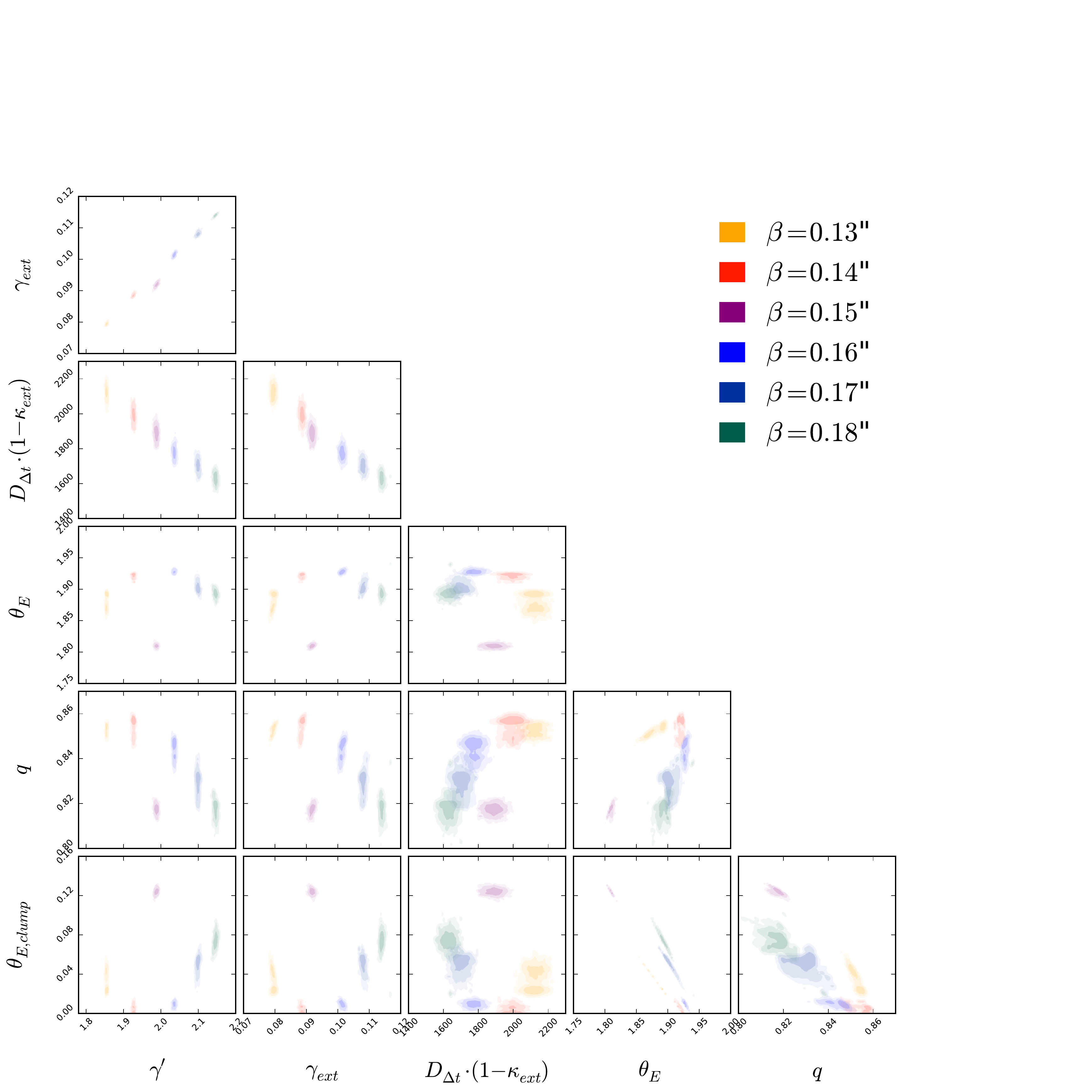}
  \caption{Posterior distribution (1-2-3 sigma contours) of lens model parameters and time delay distance of the combined analysis of imaging data of F555W and time delay measurements. Different colors correspond to different choices of the shapelet scale $\beta$. (same as Figure \ref{fig:RXJ1131_constraints} for filter F814W).}
\label{fig:RXJ1131_F555W_constraints}
\end{figure*}

In the paper, we did focus on the analysis of the WFC1 F814W filter band. Here we present the same analysis for filter F555W. Figure \ref{fig:RXJ1131_F555W_constraints} shows the posterior distribution of the lens model parameters and time delay distance for F555W. Figure \ref{fig:RXJ1131_F555W_Dd_Ds_Dsd} shows the constraints on the angular diameter distance relation. The values describing the distribution can be found in the main text.

\section{Bayesian description and renormalization of the imaging likelihood} \label{app:renormalization_bayes}
One of the steps presented in this paper is the renormalization of the imaging likelihood for different source scales $\beta$. In Section \ref{sec:relax_lens_model} we provided heuristic arguments for this approach in the case of time delay cosmography. In the following Section, we provide a Bayesian interpretation and justification of our choice in performing this calculation.

Let us assume that there is a complete model that is able to fully describe the lens, with parameters $\alpha$. However, when we fit the data, in our modeling process, we use a restricted subset of the model containing only the parameters $\hat{\alpha}$ and that the missing degrees of freedom are captured by the parameters $\theta$. To complete our notations, the source scale is given as $\beta$, the cosmological parameters as $\pi$. We also denote the image data as $D_{I}$, the kinematic data as $D_{\sigma}$ and any other independent data of the time delays and the lens environment as $D_{o}$.

Our goal is to estimate the cosmological parameters $\pi$ given the data, which is $P(\pi| D_I, D_{\sigma}, D_{o})$. We can state, using Bayes rule

\begin{equation}
  P(\pi| D_I, D_{\sigma}, D_{o}) = \int P(\pi| \hat{\alpha}, \theta, \beta, D_I, D_{\sigma}, D_{o}) P(\hat{\alpha}, \theta, \beta | D_I, D_{\sigma}, D_{o})d\hat{\alpha} d\theta d\beta.
\end{equation}
Independence of $D_I$, $D_{\sigma}$ and $D_{o}$ results in
\begin{equation}
  P(\pi| D_I, D_{\sigma}, D_{o}) = \int P(\pi| \hat{\alpha}, \theta, \beta, D_I, D_{\sigma}, D_{o}) P(\hat{\alpha}, \theta, \beta | D_I) P(\hat{\alpha}, \theta, \beta | D_{\sigma}, D_{o})d\hat{\alpha} d\theta d\beta.
\end{equation}
The internal part of the MST is encapsulated in the term $P(\hat{\alpha}, \theta, \beta | D_I)$. One way to think about MST is that the source scale cannot be measured from imaging data alone. In other words, given image data and marginalizing over all possible lens models, one should recover the source size prior. The Bayesian expression for the MST is then 
\begin{equation}
  \int P(\hat{\alpha}, \theta, \beta | D_I) d\hat{\alpha} d\theta = P(\beta),
\end{equation}
which can also be written as
\begin{equation}
  P(\hat{\alpha}, \theta, \beta | D_I) = P(\hat{\alpha}, \theta | D_I, \beta) P(\beta | D_I) = P(\hat{\alpha}, \theta | D_I, \beta) P(\beta).
\end{equation}
Incorporating this into the earlier expression we get
\begin{equation}
  P(\pi| D_I, D_{\sigma}, D_{o}) = \int P(\pi| \hat{\alpha}, \theta, \beta, D_I, D_{\sigma}, D_{o}) P(\hat{\alpha}, \theta | D_I, \beta) P(\beta) P(\hat{\alpha}, \theta, \beta | D_{\sigma}, D_{o})d\hat{\alpha} d\theta d\beta.
\end{equation}
This can be simplified further by considering the dependencies of the variables. For instance $P(\pi| \hat{\alpha}, \theta, \beta, D_I, D_{\sigma}, D_{o})$ simplifies to $P(\pi| \hat{\alpha}, \theta, D_{\sigma}, D_{o})$, since all the information from $D_I$ and $\beta$ are captured by $\hat{\alpha}$ and $\theta $. Further more the parameter $\beta$ is not directly dependent on the velocity dispersion $D_{\sigma}$ and related quantities $D_{o}$ through the lens model. This relation of parameters and conditional data leads to
\begin{equation}
  P(\pi| D_I, D_{\sigma}, D_{o}) = \int P(\pi| \hat{\alpha}, \theta, D_{\sigma}, D_{o}) P(\hat{\alpha}, \theta | D_I, \beta) P(\beta) P(\hat{\alpha}, \theta | D_{\sigma}, D_{o})d\hat{\alpha} d\theta d\beta.
\end{equation}
Until now, no approximations are made in the Bayesian analysis. The split of $\alpha \rightarrow (\hat{\alpha}, \theta)$ has been useful in working out the impact of the internal MST in our Bayesian analysis. However, to move the analysis further, we will have to make some simplifying assumptions about the further, i.e. beyond MST, impact of the unknown lens model parameters $\theta$. \cite{Kochanek:2002p13078} showed that time delays (and hence the cosmological inference) depends mostly on the slope of the density profile in the annulus over which the lens images are observed, which is part of $\hat{\alpha}$ in our model. From this we assume that $\hat{\alpha}$ is a good approximation of the overall lens model $\alpha$ and the relative deviation $\theta$ is small in terms of the impact on the cosmographic analysis ($\pi$). We approximate $\alpha \approx \hat{\alpha}$ at this stage, which leads to

\begin{equation}
  P(\pi| D_I, D_{\sigma}, D_{o}) \approx \int P(\pi| \hat{\alpha}, D_{o}) P(\hat{\alpha} | D_I, \beta) P(\beta) P(\hat{\alpha} | D_{\sigma}, D_{o})d\hat{\alpha} d\beta.
\end{equation}
This equation is the formal expression of the steps that we perform in our analysis of combining imaging, time-delay, kinematic and environment data in our cosmographic analysis. The imaging data $D_I$ folds in the analysis through the term $P(\hat{\alpha} | D_I, \beta)$. This term is conditional on the source scale $\beta$. This conditional likelihood is effectively computed by a renormalization of the imaging likelihood to a given source scale $\beta$.

\section{Skewed normal distribution} \label{app:skew_normal_distribution}
The the skewed normal distribution is defined with a parameter $\alpha$ as
\begin{equation}
  \phi_{\gamma}(x) = 2\phi(x)\Phi(\alpha x)
\end{equation}
with $\phi(x)$ being the standard normal probability density function and $\Phi(x)$ its cumulative distribution. Location and scale can be added with
\begin{equation}
  x \rightarrow \frac{x - \xi}{\omega}.
\end{equation}
The mean $\mu$ of this distribution is given by
\begin{equation} \label{eqn:skew_mean}
  \mu = \xi + \omega\delta\sqrt{\frac{2}{\pi}} 
\end{equation}
where
\begin{equation}
  \delta = \frac{\alpha}{\sqrt{1+\alpha^2}}.
\end{equation}
The variance $\sigma^2$ is
\begin{equation}
  \sigma^2 = \omega^2\left(1 - \frac{2\delta^2}{\pi}\right)
\end{equation}
and the skewness $\gamma$ as
\begin{equation} \label{eqn:skew_skewness}
  \gamma = \frac{4-\pi}{2} \frac{\left(\delta\sqrt{2/\pi}\right)^3}{  \left(1-2\delta^2/\pi\right)^{3/2}}.
\end{equation}
The skewed normal distribution $\phi_{\gamma}(x, \xi, \omega, \alpha)$ can be re-parameterized to $\phi_{\gamma}(x, \mu, \sigma, \gamma)$ by inverting the equations (\ref{eqn:skew_mean})-(\ref{eqn:skew_skewness}).

\section{Lens model parameter constraints} \label{app:lens_model_parameter_constraints}
Figure \ref{fig:RXJ1131_constraints} and \ref{fig:RXJ1131_F555W_constraints} show a selection of posteriors of the lens model parameters for the two images in band F814W and F555W in combination with the time delay measurements. In this section, we list the one-dimensional posteriors of all the lens model parameters involved. These constraints are listed in Table \ref{tab:lens_params_F814W} for the F814W analysis and in Table \ref{tab:lens_params_F555W} for the F555W analysis.

\begin{sidewaystable}
  \caption{Lens model parameter posterior of F814W.}

  \begin{tabular}{lrrrrrr}
  \hline \hline
  Parameter & $\beta = 0.14$ & $\beta = 0.15$ & $\beta = 0.16$ & $\beta = 0.17$ & $\beta = 0.18$ & $\beta = 0.19$ \\
  \hline
$\theta_E$  &  $ 1.7997  ^{+ 0.0017 }_{- 0.0016 }$  &  $ 1.7681  ^{+ 0.0011 }_{- 0.0013 }$  &  $ 1.9202  ^{+ 0.001 }_{- 0.0011 }$  &  $ 1.8452  ^{+ 0.0012 }_{- 0.001 }$  &  $ 1.9048  ^{+ 0.0005 }_{- 0.0005 }$  &  $ 1.832  ^{+ 0.0015 }_{- 0.0015 }$  \\

$\gamma'$ &  $ 1.8697  ^{+ 0.0014 }_{- 0.0013 }$  &  $ 1.8679  ^{+ 0.0017 }_{- 0.0017 }$  &  $ 2.0114  ^{+ 0.0016 }_{- 0.0016 }$  &  $ 2.0787  ^{+ 0.0018 }_{- 0.0019 }$  &  $ 2.1356  ^{+ 0.0019 }_{- 0.0019 }$  &  $ 2.1936  ^{+ 0.0025 }_{- 0.0024 }$  \\  
$x_{0, \text{lens}}$ &  $ 0.1618  ^{+ 0.0003 }_{- 0.0004 }$  &  $ 0.17  ^{+ 0.0014 }_{- 0.001 }$  &  $ 0.2168  ^{+ 0.0004 }_{- 0.0005 }$  &  $ 0.1924  ^{+ 0.0006 }_{- 0.0006 }$  &  $ 0.2144  ^{+ 0.0004 }_{- 0.0004 }$  &  $ 0.1954  ^{+ 0.0025 }_{- 0.0027 }$  \\
$y_{0, \text{lens}}$ & $ -0.1549  ^{+ 0.0004 }_{- 0.0004 }$  &  $ -0.1502  ^{+ 0.0006 }_{- 0.0006 }$  &  $ -0.1257  ^{+ 0.0003 }_{- 0.0003 }$  &  $ -0.1466  ^{+ 0.0004 }_{- 0.0004 }$  &  $ -0.1414  ^{+ 0.0003 }_{- 0.0003 }$  &  $ -0.1642  ^{+ 0.0006 }_{- 0.0006 }$  \\
$q$ &  $ 0.8435  ^{+ 0.0006 }_{- 0.0006 }$  &  $ 0.8356  ^{+ 0.0007 }_{- 0.0006 }$  &  $ 0.8765  ^{+ 0.0008 }_{- 0.0007 }$  &  $ 0.844  ^{+ 0.001 }_{- 0.001 }$  &  $ 0.866  ^{+ 0.0007 }_{- 0.0007 }$  &  $ 0.8239  ^{+ 0.0028 }_{- 0.0028 }$  \\
$\phi_q$ &  $ -0.426  ^{+ 0.0008 }_{- 0.0008 }$  &  $ -0.4442  ^{+ 0.0009 }_{- 0.001 }$  &  $ -0.51  ^{+ 0.0028 }_{- 0.0026 }$  &  $ -0.4506  ^{+ 0.0016 }_{- 0.0016 }$  &  $ -0.4819  ^{+ 0.0017 }_{- 0.0016 }$  &  $ -0.4251  ^{+ 0.0031 }_{- 0.0031 }$  \\ 
$x_{0, \text{qso}}$ &  $ 0.5881  ^{+ 0.0007 }_{- 0.0007 }$  &  $ 0.592  ^{+ 0.001 }_{- 0.001 }$  &  $ 0.6735  ^{+ 0.0008 }_{- 0.0009 }$  &  $ 0.7006  ^{+ 0.0009 }_{- 0.0009 }$  &  $ 0.7305  ^{+ 0.0009 }_{- 0.0009 }$  &  $ 0.7499  ^{+ 0.001 }_{- 0.001 }$  \\  
$y_{0, \text{qso}}$ & $ -0.2082  ^{+ 0.0003 }_{- 0.0003 }$  &  $ -0.1888  ^{+ 0.0004 }_{- 0.0004 }$  &  $ -0.2073  ^{+ 0.0004 }_{- 0.0004 }$  &  $ -0.2112  ^{+ 0.0004 }_{- 0.0004 }$  &  $ -0.2231  ^{+ 0.0004 }_{- 0.0004 }$  &  $ -0.2229  ^{+ 0.0005 }_{- 0.0005 }$  \\
$\theta_{E,\text{clump}}$ &  $ 0.1098  ^{+ 0.0015 }_{- 0.0016 }$  &  $ 0.1356  ^{+ 0.0012 }_{- 0.001 }$  &  $ 0.0119  ^{+ 0.0011 }_{- 0.0011 }$  &  $ 0.1092  ^{+ 0.0014 }_{- 0.0015 }$  &  $ 0.0593  ^{+ 0.0008 }_{- 0.0009 }$  &  $ 0.1645  ^{+ 0.0022 }_{- 0.0024 }$  \\
$\gamma_{\text{ext}}$ &  $ 0.0722  ^{+ 0.0002 }_{- 0.0002 }$  &  $ 0.0726  ^{+ 0.0003 }_{- 0.0003 }$  &  $ 0.0964  ^{+ 0.0003 }_{- 0.0003 }$  &  $ 0.1019  ^{+ 0.0003 }_{- 0.0003 }$  &  $ 0.1115  ^{+ 0.0003 }_{- 0.0003 }$  &  $ 0.1127  ^{+ 0.0004 }_{- 0.0004 }$  \\ 
$\phi_{\text{ext}}$ &  $ 0.7259  ^{+ 0.0007 }_{- 0.0007 }$  &  $ 0.7737  ^{+ 0.0012 }_{- 0.0012 }$  &  $ 0.7055  ^{+ 0.0007 }_{- 0.0007 }$  &  $ 0.6932  ^{+ 0.0007 }_{- 0.0007 }$  &  $ 0.6727  ^{+ 0.0005 }_{- 0.0006 }$  &  $ 0.6687  ^{+ 0.0008 }_{- 0.0008 }$  \\ 
$D_{\Delta t} \cdot (1-\kappa_{\text{ext}})$ &  $ 2150^{+ 36}_{- 39}$  &  $ 2108^{+ 32}_{- 32}$  &  $ 1851 ^{+ 30}_{- 28}$  &  $ 1775  ^{+ 29}_{-28}$  &  $ 1695^{+ 26}_{- 27}$  &  $ 1644  ^{+ 28 }_{- 28 }$  \\
  \hline \hline

  \end{tabular}
  \label{tab:lens_params_F814W}
\end{sidewaystable}

\begin{sidewaystable}
  \caption{Lens model parameter posterior of F555W.}

  \begin{tabular}{lrrrrrr}
  \hline \hline
  Parameter & $\beta = 0.14$ & $\beta = 0.15$ & $\beta = 0.16$ & $\beta = 0.17$ & $\beta = 0.18$ & $\beta = 0.19$ \\
  \hline

$\theta_E$  &  $ 1.881  ^{+ 0.012 }_{- 0.014 }$  &  $ 1.922  ^{+ 0.003 }_{- 0.004 }$  &  $ 1.81  ^{+ 0.002 }_{- 0.002 }$  &  $ 1.928  ^{+ 0.002 }_{- 0.002 }$  &  $ 1.902  ^{+ 0.007 }_{- 0.006 }$  &  $ 1.894  ^{+ 0.005 }_{- 0.005 }$  \\  $\gamma'$  &  $ 1.855  ^{+ 0.002 }_{- 0.002 }$  &  $ 1.927  ^{+ 0.002 }_{- 0.002 }$  &  $ 1.988  ^{+ 0.003 }_{- 0.003 }$  &  $ 2.036  ^{+ 0.003 }_{- 0.003 }$  &  $ 2.099  ^{+ 0.003 }_{- 0.003 }$  &  $ 2.146  ^{+ 0.003 }_{- 0.003 }$  \\  $x_{0, \text{lens}}$  &  $ 0.1473  ^{+ 0.0008 }_{- 0.0009 }$  &  $ 0.1527  ^{+ 0.0045 }_{- 0.008 }$  &  $ 0.1327  ^{+ 0.0012 }_{- 0.0009 }$  &  $ 0.1614  ^{+ 0.0041 }_{- 0.0063 }$  &  $ 0.1539  ^{+ 0.0068 }_{- 0.0085 }$  &  $ 0.1586  ^{+ 0.0036 }_{- 0.005 }$  \\  $y_{0, \text{lens}}$  &  $ -0.1436  ^{+ 0.003 }_{- 0.0037 }$  &  $ -0.14  ^{+ 0.0009 }_{- 0.0013 }$  &  $ -0.1723  ^{+ 0.0008 }_{- 0.0008 }$  &  $ -0.142  ^{+ 0.0009 }_{- 0.0009 }$  &  $ -0.1553  ^{+ 0.0031 }_{- 0.003 }$  &  $ -0.1509  ^{+ 0.0018 }_{- 0.0015 }$  \\  $q$  &  $ 0.8526  ^{+ 0.0022 }_{- 0.0024 }$  &  $ 0.8554  ^{+ 0.002 }_{- 0.004 }$  &  $ 0.8175  ^{+ 0.0016 }_{- 0.0017 }$  &  $ 0.8448  ^{+ 0.003 }_{- 0.0042 }$  &  $ 0.8294  ^{+ 0.0032 }_{- 0.0046 }$  &  $ 0.8181  ^{+ 0.0022 }_{- 0.0033 }$  \\  $\phi_q$  &  $ -0.5206  ^{+ 0.0036 }_{- 0.0036 }$  &  $ -0.5172  ^{+ 0.0095 }_{- 0.0051 }$  &  $ -0.4714  ^{+ 0.0021 }_{- 0.002 }$  &  $ -0.5071  ^{+ 0.0094 }_{- 0.0068 }$  &  $ -0.4834  ^{+ 0.0077 }_{- 0.0054 }$  &  $ -0.4873  ^{+ 0.0051 }_{- 0.0033 }$  \\  $x_{0, \text{qso}}$  &  $ 0.5688  ^{+ 0.001 }_{- 0.0011 }$  &  $ 0.608  ^{+ 0.0017 }_{- 0.0019 }$  &  $ 0.6344  ^{+ 0.0017 }_{- 0.0018 }$  &  $ 0.6673  ^{+ 0.0015 }_{- 0.0015 }$  &  $ 0.6976  ^{+ 0.0015 }_{- 0.0015 }$  &  $ 0.7234  ^{+ 0.0016 }_{- 0.0016 }$  \\  $y_{0, \text{qso}}$  &  $ -0.2124  ^{+ 0.0006 }_{- 0.0006 }$  &  $ -0.2261  ^{+ 0.0007 }_{- 0.0007 }$  &  $ -0.2326  ^{+ 0.0007 }_{- 0.0007 }$  &  $ -0.235  ^{+ 0.0006 }_{- 0.0006 }$  &  $ -0.2458  ^{+ 0.0007 }_{- 0.0007 }$  &  $ -0.2333  ^{+ 0.0008 }_{- 0.0007 }$  \\  $\theta_{E,\text{clump}}$  &  $ 0.0332  ^{+ 0.0121 }_{- 0.0102 }$  &  $ 0.003  ^{+ 0.0037 }_{- 0.0028 }$  &  $ 0.124  ^{+ 0.0023 }_{- 0.0023 }$  &  $ 0.0094  ^{+ 0.0025 }_{- 0.0025 }$  &  $ 0.051  ^{+ 0.006 }_{- 0.0066 }$  &  $ 0.073  ^{+ 0.0062 }_{- 0.0054 }$  \\  $\gamma_{\text{ext}}$  &  $ 0.0794  ^{+ 0.0005 }_{- 0.0005 }$  &  $ 0.0886  ^{+ 0.0004 }_{- 0.0005 }$  &  $ 0.0918  ^{+ 0.0006 }_{- 0.0005 }$  &  $ 0.1014  ^{+ 0.0005 }_{- 0.0006 }$  &  $ 0.108  ^{+ 0.0005 }_{- 0.0005 }$  &  $ 0.114  ^{+ 0.0004 }_{- 0.0004 }$  \\  $\phi_{\text{ext}}$  &  $ 0.7779  ^{+ 0.0021 }_{- 0.0023 }$  &  $ 0.7384  ^{+ 0.0013 }_{- 0.0014 }$  &  $ 0.7296  ^{+ 0.0015 }_{- 0.0015 }$  &  $ 0.7169  ^{+ 0.0012 }_{- 0.0012 }$  &  $ 0.6966  ^{+ 0.0011 }_{- 0.0011 }$  &  $ 0.7119  ^{+ 0.0011 }_{- 0.0011 }$  \\  $D_{\Delta t} \cdot (1-\kappa_{\text{ext}})$  &  $ 2124.0  ^{+ 33.0 }_{- 35.0 }$  &  $ 1995.0  ^{+ 32.0 }_{- 33.0 }$  &  $ 1888.0  ^{+ 30.0 }_{- 31.0 }$  &  $ 1775.0  ^{+ 29.0 }_{- 29.0 }$  &  $ 1701.0  ^{+ 29.0 }_{- 28.0 }$  &  $ 1630.0  ^{+ 27.0 }_{- 27.0 }$  \\
  \hline \hline

  \end{tabular}
  \label{tab:lens_params_F555W}
\end{sidewaystable}

\section{Source size prior} \label{app:source_size_prior}
To account for an arbitrary prior in $\beta$ in the Bayesian inference, one has to marginalize as
\begin{equation} \label{eqn:prior_likelihood}
  P(\boldsymbol{d_{\text{RXJ}}} | \boldsymbol{\pi}) = \int P(\boldsymbol{d_{\text{RXJ}}}| \boldsymbol{\boldsymbol{\pi}}, \beta)P(\beta)d\beta = \int P(\boldsymbol{d_{\text{RXJ}}}| \boldsymbol{\pi})P(\boldsymbol{\pi}|\beta)P(\beta)d\beta.
\end{equation}
$\beta$ does not appear as a parameter in the likelihood of Equation \ref{eqn:skew_likelihood}. From Figure \ref{fig:RXJ1131_Dd_Ds_Dsd} one sees that the different source scale posteriors are equally spaced in the $D_{ds}/(D_dD_s)$ axis. The likelihood defined in Equation \ref{eqn:skew_likelihood} is an approximation for a flat prior in the source scale $\beta$. We approximate $P(\boldsymbol{\pi}|\beta)$ by a delta function in the parameter $D_{ds}/(D_dD_s)$ as
\begin{equation}
  P(\boldsymbol{\pi}|\beta) \approx \delta \left(\alpha_{\beta} \frac{D_{ds}}{D_dD_s} + C_{\beta} - \beta \right)
\end{equation}
where $\alpha_{\beta}$ is the slope of the $D_{ds}/(D_dD_s)$ vs $\beta$ and $C_{\beta}$ the intercept. In this form, the prior on $\beta$ can be added to the likelihood of Equation \ref{eqn:skew_likelihood} as
\begin{equation} \label{eqn:source_size_prior}
  P(\boldsymbol{d_{\text{RXJ}}}, \boldsymbol{\pi}) = \phi_{\gamma} \left( x=\frac{D_{\text{ds}}}{D_{\text{d}} D_{\text{s}}}, \mu = \alpha \ln (D_{\text{d}}) + C, \sigma_D, \gamma_D \right) \left(\alpha_{\beta} \frac{D_{ds}}{D_dD_s} + C_{\beta}\right)^{2\alpha_{\text{LF}}+1}.
\end{equation}

\end{document}